\documentclass[nobibnotes,superscriptaddress,nofootinbib,twocolumn]{revtex4-1}
\usepackage{amsmath,amssymb,multirow}
\usepackage{graphicx}

\begin{document}

\title{Maximally restrictive leptonic texture zeros in two-Higgs-doublet models}

\author{R.~Gonz\'{a}lez Felipe}
\email{ricardo.felipe@tecnico.ulisboa.pt}
\affiliation{ISEL - Instituto Superior de Engenharia de Lisboa, Instituto Polit\'ecnico de Lisboa, Rua Conselheiro Em\'{\i}dio Navarro 1959-007 Lisboa,	Portugal}
\affiliation{Departamento de F\'{\i}sica and Centro de F\'{\i}sica Te\'{o}rica de Part\'{\i}culas - CFTP, Instituto Superior T\'{e}cnico, Universidade de Lisboa, Av. Rovisco Pais, 1049-001 Lisboa,
	Portugal}

\author{H. Ser\^{o}dio}
\email{hugo.serodio@thep.lu.se}
\affiliation{Department of Physics, Korea University, Seoul 136-713, Korea}
\affiliation{Department of Astronomy and Theoretical Physics, Lund University, SE-223 62 Lund, Sweden}
	

\begin{abstract}
The implementation of maximally restrictive texture zeros in the leptonic sector is investigated in the context of two-Higgs-doublet models with Majorana neutrinos. After analyzing all maximally restrictive pairs of leptonic mass matrices with zero entries, we conclude that there are only four texture combinations that are compatible with observations at $3\sigma$ confidence level and can be implemented through Abelian symmetries in a two-Higgs-doublet model. The compatibility of these textures with current constraints on lepton-flavor-violating processes is also studied. The ultraviolet completion of these models is discussed in the framework of the seesaw mechanism for neutrino masses.
\end{abstract}

\maketitle

\section{Introduction}
\label{sec:intro}

The existence of neutrino masses and leptonic mixing has been firmly established by neutrino oscillation experiments~\cite{Fogli:2012ua,GonzalezGarcia:2012sz,Forero:2014bxa}. However there is no convincing theory to explain their origin and several questions remain to be answered~\cite{Strumia:2006db,Nunokawa:2007qh,Branco:2011zb}. In particular, neutrinos can be Dirac or Majorana particles, and the CP symmetry can be violated in the lepton sector. Furthermore, the absolute neutrino mass scale and type of mass spectrum (normal or inverted hierarchy) are still unknown.

A common approach towards the solution of the flavor puzzle consists on requiring some of the  elements in the leptonic mass matrices to vanish, leading to the so-called texture zeros. It is well known that such zeros can be enforced in arbitrary entries of the fermion mass matrices by means of Abelian symmetries~\cite{Grimus:2004hf}. Moreover, combining the canonical and Smith normal form methods~\cite{Ivanov:2011ae,Serodio:2013gka,Ivanov:2013bka}, it is possible to construct viable patterns in a minimal framework, i.e. with the smallest discrete Abelian group and a minimal number of Higgs scalars. For instance, these two methods have been successfully employed to construct the minimal Abelian symmetry realizations of the well-known Frampton-Glashow-Marfatia two-zero neutrino textures~\cite{Frampton:2002yf} and to study their implementation in extensions of the standard model (SM) based on the seesaw mechanism for the neutrino masses~\cite{Felipe:2014vka}.

Recently, a complete survey of texture zeros in the lepton mass matrices has been performed~\cite{Ludl:2014axa}. Admitting all possible combinations of patterns with zeros, the authors of Ref.~\cite{Ludl:2014axa} concluded that several classes of texture zeros in the charged-lepton and neutrino mass matrices are compatible with current neutrino oscillation data at  $5\sigma$ confidence level (CL). In this paper, we reconsider this problem with the aim of identifying the maximally restrictive pairs of leptonic mass matrices with zero entries that are not only compatible with observations but also realizable in terms of Abelian symmetries. In this context, by maximally restrictive we mean that no additional zero can be placed into one of the two lepton mass matrices while keeping compatibility with the observed lepton masses and mixing data. Since none of the textures can be implemented through such symmetries in the SM, we shall consider one of its simplest extensions, namely the two-Higgs-doublet model (2HDM)~\cite{Branco:2011iw}. We also assume that neutrinos are Majorana particles. The ultraviolet (UV) completion of these models is then implemented in the framework of the seesaw mechanism for neutrino masses.

\section{Confronting texture zeros with neutrino data}
\label{sec:confront}

A detailed analysis of texture zeros in the charged-lepton mass matrix $m_\ell$ and the effective Majorana neutrino mass matrix $m_\nu$ has been carried out in Ref.~\cite{Ludl:2014axa}. In particular, it has been shown that, without imposing any correlation among the mass matrix elements, at least eight physical parameters\footnote{The number of physical parameters corresponds to the number of real parameters in the leptonic mass textures that remain after removing all unphysical phases through weak-basis transformations.} are required to explain the eight experimentally known observables (three charged-lepton masses, two neutrino mass-squared differences and three mixing angles). Moreover, several combinations of texture zeros in the pairs of mass matrices $(m_\ell,m_\nu)$ were found viable at $5\sigma$ CL~(see~\cite{Ludl:2014axa} for further details).

In this section, we shall look for $(m_\ell,m_\nu)$ pairs that contain maximally restrictive texture zeros and, simultaneously, are compatible with observations at $3\sigma$ CL. We shall perform a $\chi^2$-analysis using the standard function
\begin{equation}\label{chisquared}
\chi^2(x) = \sum_{i}
\frac{(\mathcal{P}_i(x)-\overline{\mathcal{O}}_i)^2}{\sigma_i^2},
\end{equation}
where $x$ denotes the physical input parameters (i.e. the matrix elements of $m_\ell$ and $m_\nu$), $\mathcal{P}_i(x)$ are the predictions of the model for the observables $\mathcal{O}_i$, $\overline{\mathcal{O}}_i$ are the best-fit values of $\mathcal{O}_i$, and $\sigma_i$ their corresponding $1\sigma$ errors. We use the current global fit of neutrino oscillation data at $3\sigma$~\cite{Forero:2014bxa}. We also impose the recent cosmological constraint on the sum of the neutrino masses, namely $\sum_i m_i < 0.12~\text{eV}$ ($i=1,2,3$) at 95\% CL~\cite{Palanque-Delabrouille:2015pga} (see also Ref.~\cite{Hannestad:2016fog,Alam:2016hwk}). 

In our search for a viable charged-lepton mass matrix $m_\ell$, we require that the eigenvalues of the input matrix correctly reproduce the central values of the charged-lepton masses~\cite{Agashe:2014kda}. The $\chi^2$-function is then minimized with respect to the five neutrino observables (two neutrino mass-squared differences and three mixing angles) using MINUIT~\cite{James:1975dr} and the numerical strategy presented in Ref.~\cite{Cebola:2015dwa}. Since current neutrino data does not constrain the Dirac CP-violating phase $\delta$ at $3\sigma$ (i.e. $\delta$ is allowed to vary between 0 and $2\pi$), we do not fit it in our analysis. If the deviation of each neutrino observable from its experimental value is at most $3\sigma$ at the minimum of $\chi^2$ for a given $(m_\ell,m_\nu)$-pair, the corresponding leptonic textures are said to be compatible with data.

From our $\chi^2$-analysis, several charged-lepton and Majorana neutrino mass matrices resulted in pairs of maximally restrictive textures (i.e. with 8 physical parameters) compatible with data at $3\sigma$ CL. These patterns are presented in Tables~\ref{table1} and \ref{table2}. Note that we have kept the notation of Ref.~\cite{Ludl:2014axa} for the labeling of the textures. The corresponding minimum of $\chi^2$ for the viable pairs of leptonic textures is given in Table~\ref{table3}. There are 19 pairs of leptonic matrices that can successfully accommodate the data. The lowest value of $\chi^2$ was attained for the matrix combinations $(\mathbf{5_1^{\ell}},\mathbf{3_3^{\nu}})$ and $(\mathbf{5_1^{\ell}},\mathbf{3_7^{\nu}})$. In all cases, the neutrino mass spectrum has an inverted ordering, i.e. $m_3<m_1<m_2$. Moreover, the six viable four-zero neutrino textures, $\mathbf{4_1^{\nu}}, \ldots,\mathbf{4_6^{\nu}}$, contain a null row and a null column, thus leading to a massless neutrino ($m_3=0$). 

\begin{table}[t]
	\centering
\begin{tabular}{ll}
	\hline\\[-0.3cm]
	$\mathbf{4_1^{\ell}}\sim$
	$\begin{pmatrix}
	0 & 0 & \times \\
	0 & \times & 0 \\
	\times & \times & \times\end{pmatrix}$
	&
	$\mathbf{4_3^{\ell}}\sim$
	$\begin{pmatrix}
	0 & 0 & \times \\
	0 & \times & \times \\
	\times & \times & 0\end{pmatrix}$
	\\[0.6cm]
	\hline\\[-0.3cm]
	$\mathbf{5_1^{\ell}}\sim$
	$\begin{pmatrix}
	0 & 0 & \times \\
	0 & \times & 0 \\
	\times & 0 & \times\end{pmatrix}$
	&
	\\[0.6cm]
	\hline
\end{tabular}
\caption{Viable maximally restrictive charged-lepton mass matrices.}
\label{table1}
\end{table}

\begin{table}[t]
	\centering
\begin{tabular}{ll}
	\hline\\[-0.3cm]
	$\mathbf{3_1^{\nu}}\sim$
	$\begin{pmatrix}
	0 & \times & \times \\
	\times & 0 & \times \\
	\times & \times & 0\end{pmatrix}$
	&
	$\mathbf{3_2^{\nu}}\sim$
	$\begin{pmatrix}
	0 & 0 & \times \\
	0 & 0 & \times \\
	\times & \times & \times\end{pmatrix}$
	\\
	$\mathbf{3_3^{\nu}}\sim$
	$\begin{pmatrix}
	0 & 0 & \times \\
	0 & \times & \times \\
	\times & \times & 0\end{pmatrix}$
	&
    $\mathbf{3_6^{\nu}}\sim$
	$\begin{pmatrix}
	0 & \times & \times \\
	\times & 0 & 0 \\
	\times & 0 & \times\end{pmatrix}$
	\\
	$\mathbf{3_7^{\nu}}\sim$
	$\begin{pmatrix}
	0 & \times & \times \\
	\times & \times & 0 \\
	\times & 0 & 0\end{pmatrix}$
	&
	$\mathbf{3_8^{\nu}}\sim$
	$\begin{pmatrix}
	\times & 0 & \times \\
	0 & 0 & \times \\
	\times & \times & 0\end{pmatrix}$
	\\
	\!\!$\mathbf{3_{10}^{\nu}}\sim$
	$\begin{pmatrix}
	\times & \times & \times \\
	\times & 0 & 0 \\
	\times & 0 & 0\end{pmatrix}$
	\\[0.6cm]
	\hline\\[-0.3cm]
	$\mathbf{4_1^{\nu}}\sim$
	$\begin{pmatrix}
	0 & 0 & 0 \\
	0 & 0 & \times \\
	0 & \times & \times\end{pmatrix}$
	&
	$\mathbf{4_2^{\nu}}\sim$
	$\begin{pmatrix}
	0 & 0 & 0 \\
	0 & \times & \times \\
	0 & \times & 0\end{pmatrix}$
	\\
	$\mathbf{4_3^{\nu}}\sim$
	$\begin{pmatrix}
	0 & 0 & \times \\
	0 & 0 & 0 \\
	\times & 0 & \times\end{pmatrix}$
    &
	$\mathbf{4_4^{\nu}}\sim$
	$\begin{pmatrix}
	0 & \times & 0 \\
	\times & \times & 0 \\
	0 & 0 & 0\end{pmatrix}$
	\\
	$\mathbf{4_5^{\nu}}\sim$
	$\begin{pmatrix}
	\times & 0 & \times \\
	0 & 0 & 0 \\
	\times & 0 & 0\end{pmatrix}$
	&
	$\mathbf{4_6^{\nu}}\sim$
	$\begin{pmatrix}
	\times & \times & 0 \\
	\times & 0 & 0 \\
	0 & 0 & 0\end{pmatrix}$
	\\[0.6cm]
	\hline
\end{tabular}
\caption{Viable maximally restrictive Majorana neutrino mass matrices.}
\label{table2}
\end{table}

\begin{table}[h]
	\centering
	\begin{tabular}{ccc}
		\hline
		$\mathbf{m}_\ell$ & $\mathbf{m}_\nu$ & $\chi^2_{\rm min}$\\
		\hline
		$\mathbf{4_1^{\ell}}$ & $\mathbf{4_1^{\nu}}$, $\mathbf{4_3^{\nu}}$ & 7.38\\
		                        & $\mathbf{4_4^{\nu}}$, $\mathbf{4_6^{\nu}}$ & 7.84\\
		                        & $\mathbf{4_2^{\nu}}$, $\mathbf{4_5^{\nu}}$ & 9.30\\
		\hline
	    $\mathbf{4_3^{\ell}}$ & $\mathbf{4_2^{\nu}}$,  $\mathbf{4_4^{\nu}}$ & 7.38\\
		                        & $\mathbf{4_3^{\nu}}$, $\mathbf{4_5^{\nu}}$ & 7.84\\
		                        & $\mathbf{4_1^{\nu}}$, $\mathbf{4_6^{\nu}}$ & 9.30\\
		\hline
		$\mathbf{5_1^{\ell}}$ & $\mathbf{3_3^{\nu}}$, $\mathbf{3_7^{\nu}}$ & 1.36\\
		                        & $\mathbf{3_1^{\nu}}$ & 2.04\\
                     		    & $\mathbf{3_2^{\nu}}$, $\mathbf{3_{10}^{\nu}}$ & 2.71\\
                     		    
                     		    & $\mathbf{3_6^{\nu}}$, $\mathbf{3_8^{\nu}}$ & 3.46\\
         \hline
	\end{tabular}
	\caption{The minimum of $\chi^2$ for the viable pairs of maximally restrictive leptonic textures with eight physical parameters. In all cases, the neutrino mass spectrum has an inverted ordering. Compatibility with neutrino oscillation data is found at $3\sigma$ CL.}
	\label{table3}
\end{table}

\section{Model building}
\label{sec:model}

\subsection{2HDM: the leptonic sector}

Majorana neutrino masses can be explained through the introduction of the unique dimension-five Weinberg operator compatible with the SM gauge group. Assuming two Higgs-doublet fields, the leptonic interaction Lagrangian can be written as
\begin{align}\label{Lint}
-\mathcal{L}_{\text{int}}=-\mathcal{L}_{\text{Y}}+\left[\frac{\kappa_{ij}}{2\Lambda}\,
\left(\overline{\ell_L^0}\,\tilde{\phi}_i\right)
\left(\tilde{\phi}_j^T\,\ell_{L}^{0\,c}\right)+\text{h.c.}\right],
\end{align}
where
\begin{align}\label{Lyukawa}
-\mathcal{L}_{\text{Y}}=\overline{\ell_{L}^0}\,\Pi_{i}\,
\phi_i\, e_{R}^0+\text{h.c.},
\end{align}
$\Lambda$ is an effective energy scale, $\ell_L^0$ denotes the left-handed lepton doublet fields and $e_R^0$ are the right-handed charged-lepton singlets; $\phi_i$ ($i=1,2$) are the Higgs doublets and $\tilde{\phi} \equiv i \sigma_2 \phi^\ast$; $\kappa_{ij}$ are the effective Majorana neutrino coupling matrices and $\Pi_{i}$ are the charged-lepton Yukawa coupling matrices. Note that flavor indexes have been omitted in Eqs.~\eqref{Lint} and~\eqref{Lyukawa}, and a sum over Latin indexes is assumed.

The $3\times3$ complex symmetric matrices $\kappa_{ij}$ give rise to Majorana neutrino masses once the Higgs fields
\begin{equation}
\phi_i=\frac{1}{\sqrt{2}}
\begin{pmatrix}
\sqrt{2}\, \varphi^+\\
e^{i\theta_i}(v_i+\rho_i+i\eta_i)
\end{pmatrix}
\end{equation}
acquire vacuum expectation values (vev) 
$$
\langle \phi_i \rangle \equiv \frac{1}{\sqrt{2}}
\begin{pmatrix}
0\\
e^{i\theta_i}v_i
\end{pmatrix},
$$
with $v_i$ real and positive. Without loss of generality, in what follows we set $\theta_1=0$, since only the relative phase $ \theta \equiv \theta_2-\theta_1$ is observable.

We can use a convenient basis in which an orthogonal rotation is performed so that one Higgs doublet combination has the vev $v=(v_1^2+v_2^2)^{1/2}\simeq246$~GeV and the other has none. We define
\begin{equation}
\label{eq:Higgstransf}
\begin{pmatrix}
H_1\\
H_2
\end{pmatrix}=
\begin{pmatrix}
c_\beta&s_\beta\\
-s_\beta&c_\beta
\end{pmatrix}
\begin{pmatrix}
\phi_1\\
e^{-i\theta}\phi_2
\end{pmatrix},
\end{equation}
with $c_\beta\equiv \cos\beta=v_1/v$, $s_\beta\equiv \sin\beta=v_2/v$ and
\begin{equation}
\label{eq:Higgsbasis}
H_1=\frac{1}{\sqrt{2}}
\begin{pmatrix}
\sqrt{2}\, G^+\\
v+H^0+i G^0
\end{pmatrix},\;
H_2=\frac{1}{\sqrt{2}}
\begin{pmatrix}
\sqrt{2}\, H^+\\
R+i I
\end{pmatrix}.
\end{equation}
This is known as the Higgs basis. The Yukawa Lagrangian given in Eq.~\eqref{Lyukawa} now takes the form
\begin{align}
\begin{split}
-\mathcal{L}_{\text{Y}}=& \frac{1}{v}\,\overline{e^0_{L}}\left[m_\ell(v+H^0)+N_e^0 R+iN_e^0 I\right]e^0_{R}\\
&+\frac{\sqrt{2}}{v}\,\overline{\nu^0_{L}}N_e^0 H^+e_R^0+\text{h.c.},
\end{split}
\end{align}
with 
\begin{align}
\begin{split}
m_\ell=&\frac{1}{\sqrt{2}}\left(v_1\Pi_1+e^{i\theta}v_2\Pi_2\right),\\
N^0_e=&\frac{1}{\sqrt{2}}\left(v_2\Pi_1-e^{i\theta}v_1\Pi_2\right).
\end{split}
\end{align}
The charged-lepton mass matrix $m_\ell$ is diagonalized by the bi-unitary transformation $U_{\ell L}^\dagger m_\ell \,U_{\ell R}=\text{diag}\,(m_e,m_\mu,m_\tau)$, with real and positive charged lepton masses. In this basis, $N_e=U_{\ell L}^\dagger N_e^0 \,U_{\ell R}$.  Similarly, for Majorana neutrinos, the symmetric mass matrix $m_\nu$ is diagonalized by the unitary transformation $U_{\nu L}^\dagger m_\nu\, U^{\ast}_{\nu L}=\mathrm{diag\,}(m_1,m_2,m_3)$, with $m_i$ real and positive. The leptonic mixing matrix $U$ is then given by $U=U_{\ell L}^\dagger U_{\nu L}$, which can be parametrized in terms of the three mixing angles, the Dirac phase and the two Majorana phases using the standard form~\cite{Agashe:2014kda}.

\subsection{Abelian symmetry implementation of the textures}
\label{sec:symmetry}

In this section, we study whether the maximally restrictive pairs of leptonic textures previously found (cf. Table~\ref{table3}) can accommodate an Abelian symmetry in the context of 2HDM. To achieve our goal we make use of the canonical and Smith normal form methods, following the approach presented in Ref.~\cite{Felipe:2014vka}. 

Let us first consider the charged-lepton mass matrix. There are 3 viable textures: $\mathbf{4_1^{\ell}}$, $\mathbf{4_3^{\ell}}$, and $\mathbf{5_1^{\ell}}$. Using the canonical method and looking at all possible texture decompositions, one can show that the matrix $\mathbf{4_1^{\ell}}$ cannot be obtained with just two Higgs doublets. On the other hand, the matrix $\mathbf{4_3^{\ell}}$ can be implemented via the charged-lepton Yukawa coupling textures
\begin{equation}
\label{eq:ye43}
\Pi_1=
\begin{pmatrix}
0&0&\times\\
0&\times&0\\
\times&0&0
\end{pmatrix}\,,\quad
\Pi_2=\begin{pmatrix}
0&0&0\\
0&0&\times\\
0&\times&0
\end{pmatrix},
\end{equation}
while two different decompositions can lead to the texture $\mathbf{5_1^{\ell}}$, namely 
\begin{equation}
\label{eq:ye51a}
\Pi_1=
\begin{pmatrix}
0&0&\times\\
0&\times&0\\
\times&0&0
\end{pmatrix}\,,\quad
\Pi_2=\begin{pmatrix}
0&0&0\\
0&0&0\\
0&0&\times
\end{pmatrix}\,,
\end{equation}
or
\begin{equation}
\label{eq:ye51b}
\Pi_1=
\begin{pmatrix}
0&0&\times\\
0&0&0\\
\times&0&0
\end{pmatrix}\,,\quad
\Pi_2=\begin{pmatrix}
0&0&0\\
0&\times&0\\
0&0&\times
\end{pmatrix}\,.
\end{equation}

Consider now the neutrino sector. We recall that for the Majorana neutrino mass matrix we should keep only those texture decompositions that lead to symmetric matrices. As can be seen in Table~\ref{table2}, there are 7 textures with three zeros and 6 textures with four zeros. Working in a framework with two scalar fields, $\phi_{1,2}$, we can only form three distinct combinations: $\phi_1^2$, $\phi_2^2$ and $\phi_1\phi_2$. Thus, the maximal number of different textures that build a given neutrino mass matrix is three.

Following a procedure similar to the case of charged leptons, we arrive at the following matrix decompositions for the three-zero neutrino textures:
\begin{align}
\label{eq:ynu3}
\quad\;\mathbf{3_1^{\nu}}:& 
\begin{pmatrix}
0&\times&0\\
\times& 0&0 \\
0& 0&0
\end{pmatrix},
\begin{pmatrix}
0&0&\times\\
0&0&0\\
\times&0&0
\end{pmatrix},
\begin{pmatrix}
0&0&0\\
0&0&\times\\
0&\times&0
\end{pmatrix};
\end{align}
\begin{align}
\mathbf{3_2^{\nu}}:&
\begin{pmatrix}
0&0&\times\\
0&0&0\\
\times&0&0
\end{pmatrix},
\begin{pmatrix}
0&0&0\\
0&0&\times\\
0&\times&0
\end{pmatrix},
\begin{pmatrix}
0&0&0\\
0&0&0\\
0&0&\times
\end{pmatrix}\, \text{or}\nonumber\\
&\begin{pmatrix}
0&0&\times\\
0&0&\times\\
\times&\times&0
\end{pmatrix},
\begin{pmatrix}
0&0&0\\
0&0&0\\
0&0&\times
\end{pmatrix};
\end{align}
\begin{align}
\mathbf{3_3^{\nu}}:&
\begin{pmatrix}
0&0&\times\\
0&0&0\\
\times&0&0
\end{pmatrix},
\begin{pmatrix}
0&0&0\\
0&0&\times\\
0&\times&0
\end{pmatrix},
\begin{pmatrix}
0&0&0\\
0&\times&0\\
0&0&0
\end{pmatrix}\, \text{or}\nonumber\\
&\begin{pmatrix}
0&0&\times\\
0&\times&0\\
\times&0&0
\end{pmatrix},
\begin{pmatrix}
0&0&0\\
0&0&\times\\
0&\times&0
\end{pmatrix};
\end{align}
\begin{align}
\mathbf{3_6^{\nu}}:&
\begin{pmatrix}
0&\times&0\\
\times&0&0\\
0&0&0
\end{pmatrix},
\begin{pmatrix}
0&0&\times\\
0&0&0\\
\times&0&0
\end{pmatrix},
\begin{pmatrix}
0&0&0\\
0&0&0\\
0&0&\times
\end{pmatrix}\, \text{or}\nonumber\\
&\begin{pmatrix}
0&\times&0\\
\times&0&0\\
0&0&\times
\end{pmatrix},
\begin{pmatrix}
0&0&\times\\
0&0&0\\
\times&0&0
\end{pmatrix};
\end{align}
\begin{align}
\begin{split}
\mathbf{3_7^{\nu}}:& \; P_{13}\, \mathbf{3_3^{\nu}}\, P_{13}\,;
\\
\mathbf{3_8^{\nu}}:& \; P_{13}\, \mathbf{3_6^{\nu}}\, P_{13}\,;
\\
\mathbf{3_{10}^{\nu}}:& \; P_{13}\, \mathbf{3_2^{\nu}}\, P_{13}\,.
\end{split}
\end{align}
Similarly, for the four-zero textures we obtain
\begin{align}
\label{eq:ynu4}
\mathbf{4_1^{\nu}}:&
\begin{pmatrix}
0&0&0\\
0&0&\times\\
0&\times&0
\end{pmatrix},
\begin{pmatrix}
0&0&0\\
0&0&0\\
0&0&\times
\end{pmatrix};
\end{align}
\begin{align}\label{46nu}
\begin{split}
\mathbf{4_2^{\nu}}:& \; P_{23}\, \mathbf{4_1^{\nu}}\, P_{23}\,;
\\
\mathbf{4_3^{\nu}}:& \; P_{12}\, \mathbf{4_1^{\nu}}\, P_{12}\,;
\\
\mathbf{4_4^{\nu}}:& \; P_{321}\, \mathbf{4_1^{\nu}}\, P_{321}\,;
\\
\mathbf{4_5^{\nu}}:& \; P_{123}\, \mathbf{4_1^{\nu}}\, P_{123}\,;
\\
\mathbf{4_6^{\nu}}:& \; P_{13}\, \mathbf{4_1^{\nu}}\, P_{13}\,.
\\
\end{split}
\end{align}
In the above equations, $P_{ij}$ and $P_{ijk}$ denote the $3\times3$ permutation matrices, where the indexes indicate the permutation of columns with respect to the diagonal unit matrix (see e.g.~\cite{Felipe:2014vka} for the explicit form of these matrices).

The next step is to study the possibility of realizing the above texture decompositions through Abelian symmetries. This is done more conveniently with the help of the Smith normal form method~\cite{Felipe:2014vka}. Our results are summarized in Table~\ref{table4}, where the realizable mass matrix pairs, and the corresponding Higgs combinations and Abelian symmetry are presented.  For each implementation, the coupling of $\phi_1^{\ast 2}$, $\phi_2^{\ast 2}$ and $\phi_1^\ast\phi^\ast_2$ to a given neutrino mass texture is simply denoted as the Higgs combination $(1,1)$, $(2,2)$, and $(1,2)$, respectively. We conclude from the table that only the matrix pairs $\big(\mathbf{4_3^{\ell},\mathbf{4_1^{\nu}}}\big)$,  $\big(\mathbf{4_3^{\ell},\mathbf{4_6^{\nu}}}\big)$, 
$\big(\mathbf{5_1^{\ell},\mathbf{3_6^{\nu}}}\big)$, and 	$\big(\mathbf{5_1^{\ell},\mathbf{3_8^{\nu}}}\big)$ can be implemented through an Abelian symmetry. In the case of the texture $\mathbf{5_1^{\ell}}$, the corresponding Higgs combination can be constructed using either the decomposition of Eq.~\eqref{eq:ye51a} or the one given in Eq.~\eqref{eq:ye51b}.

\begin{table}[t]
	\begin{center}
		\begin{tabular}{ccc}
			\hline\\[-0.3cm]
			$(\,m_\ell\,,\,m_\nu\,)$&Higgs combination&Symmetry\\[0.1cm]
			\hline\\[-0.3cm]
			$\big(\mathbf{4_3^{\ell},\mathbf{4_1^{\nu}}}\big)$&[(2,2),(1,2)] &\\[0.1cm]
			$\big(\mathbf{4_3^{\ell},\mathbf{4_6^{\nu}}}\big)$&[(1,1),(1,2)]  &$U(1)$\\[0.1cm]
			$\big(\mathbf{5_1^{\ell},\mathbf{3_6^{\nu}}}\big)$& [(1,1),(2,2),(1,2)]&\\[0.1cm]
	$\big(\mathbf{5_1^{\ell},\mathbf{3_8^{\nu}}}\big)$& [(2,2),(1,1),(1,2)]&\\[0.1cm]
	\hline
\end{tabular}
\end{center}
\caption{\label{table4} Viable maximally restrictive $(m_\ell,m_\nu)$-pairs and the corresponding Higgs doublet combinations and Abelian symmetry. The texture $\mathbf{5_1^{\ell}}$ can be constructed using either the decomposition of Eq.~\eqref{eq:ye51a} or the one given in Eq.~\eqref{eq:ye51b}.}
\end{table}

For completeness, next we present the charges associated to each Abelian symmetry implementation. The two viable models based on the texture $\mathbf{4_3^{\ell}}$ for the charged-lepton mass matrix share the same left-handed field transformation, 
\begin{equation}\label{symL1}
\ell^0_{L}\rightarrow \text{diag}(1,\, e^{-i2\gamma},\, e^{-i4\gamma})\,\ell^0_{L}\,,
\end{equation}
where $\gamma$ denotes the $U(1)$ global charge.
The two possible textures in the neutrino sector are then obtained through the following charge assignments:
\begin{subequations}
\begin{align}\label{symR1a}
\big(\mathbf{4_3^{\ell},\mathbf{4_1^{\nu}}}\big):\,&\left\{
\begin{array}{l}
e^0_{R}\rightarrow \text{diag}(e^{-i9\gamma},\, e^{-i7\gamma},\, e^{-i5\gamma})\,e^0_{R}\,,\\\\
\phi_1\rightarrow e^{i5\gamma}\, \phi_1\,,\quad
\phi_2 \rightarrow e^{i3\gamma}\, \phi_2\,,
\end{array}
\right.\\
&\nonumber\\
\big(\mathbf{4_3^{\ell},\mathbf{4_6^{\nu}}}\big):\,&\left\{
\begin{array}{l}
e^0_{R}\rightarrow \text{diag}(e^{-i5\gamma},\, e^{-i3\gamma},\, e^{-i\gamma})\,e^0_{R}\,,\\\\
\phi_1\rightarrow e^{i\gamma}\, \phi_1\,,\quad
\phi_2 \rightarrow e^{-i\gamma}\, \phi_2\,.
\end{array}
\right.\label{symR1b}
\end{align}
\end{subequations}

When the charged-lepton mass matrix is given by the texture $\mathbf{5_1^{\ell}}$, there are two possible realizations through Eqs.~\eqref{eq:ye51a} and~\eqref{eq:ye51b}, which we shall denote by $\mathbf{5_{1a}^{\ell}}$ and $\mathbf{5_{1b}^{\ell}}$, respectively. Both share the same left-handed field and scalar symmetry transformations. In the case of the matrix pair $\big(\mathbf{5_1^{\ell},\mathbf{3_6^{\nu}}}\big)$, these transformations are
\begin{equation}
\begin{array}{l}
\ell^0_{L}\rightarrow \text{diag}(1,\, e^{-i6\gamma},\, e^{-i2\gamma})\,\ell^0_{L}\,,\\\\
\phi_1\rightarrow e^{i3\gamma}\, \phi_1\,,\quad
\phi_2 \rightarrow e^{i\gamma}\, \phi_2\,.
\end{array}
\end{equation}
The right-handed field transformations will then dictate the implementation:
\begin{subequations}
	\begin{align}
	\big(\mathbf{5_{1a}^{\ell},\mathbf{3_6^{\nu}}}\big):\,&\quad
	e^0_{R}\rightarrow \text{diag}(e^{-i5\gamma},\, e^{-i9\gamma},\, e^{-i3\gamma})\,e^0_{R}\,,\\
	&\nonumber\\
	\big(\mathbf{5_{1b}^{\ell},\mathbf{3_6^{\nu}}}\big):\,&\quad
	e^0_{R}\rightarrow \text{diag}(e^{-i5\gamma},\, e^{-i7\gamma},\, e^{-i3\gamma})\,e^0_{R}\,.
	\end{align}
\end{subequations}
Similarly, for the matrix pair $\big(\mathbf{5_1^{\ell},\mathbf{3_8^{\nu}}}\big)$ we find
\begin{equation}
\begin{array}{l}
\ell^0_{L}\rightarrow \text{diag}(1,\, e^{i4\gamma},\, e^{-i2\gamma})\,\ell^0_{L}\,,\\\\
\phi_1\rightarrow e^{i\gamma}\, \phi_1\,,\quad
\phi_2 \rightarrow e^{-i\gamma}\, \phi_2\,,
\end{array}
\end{equation}
and
\begin{subequations}
\begin{align}
\big(\mathbf{5_{1a}^{\ell},\mathbf{3_8^{\nu}}}\big):\,&\quad
e^0_{R}\rightarrow \text{diag}(e^{-i3\gamma},\, e^{i3\gamma},\, e^{-i\gamma})\,e^0_{R}\,,\\
&\nonumber\\
\big(\mathbf{5_{1b}^{\ell},\mathbf{3_8^{\nu}}}\big):\,&\quad
e^0_{R}\rightarrow \text{diag}(e^{-i3\gamma},\, e^{i5\gamma},\, e^{-i\gamma})\,e^0_{R}\,.
\end{align}
\end{subequations}

The continuous Abelian symmetry may be discretized by setting $\gamma=2\pi/n$, with $n$ the order of the discrete $Z_n$ group. The discretization can be useful when looking at ultraviolet completions of the effective Weinberg operator, since the additional field content may allow for terms distinguishing continuous and discrete transformations. It is then useful to know what would be the minimal discrete symmetry that implement these textures. If we request that the discrete symmetry allows just these textures and no additional ones, we find that $Z_7$ is the smallest group in the six cases discussed above.

\begin{figure*}[t]
	\centering
	\begin{tabular}{l}
		\includegraphics[width=.4\textwidth]{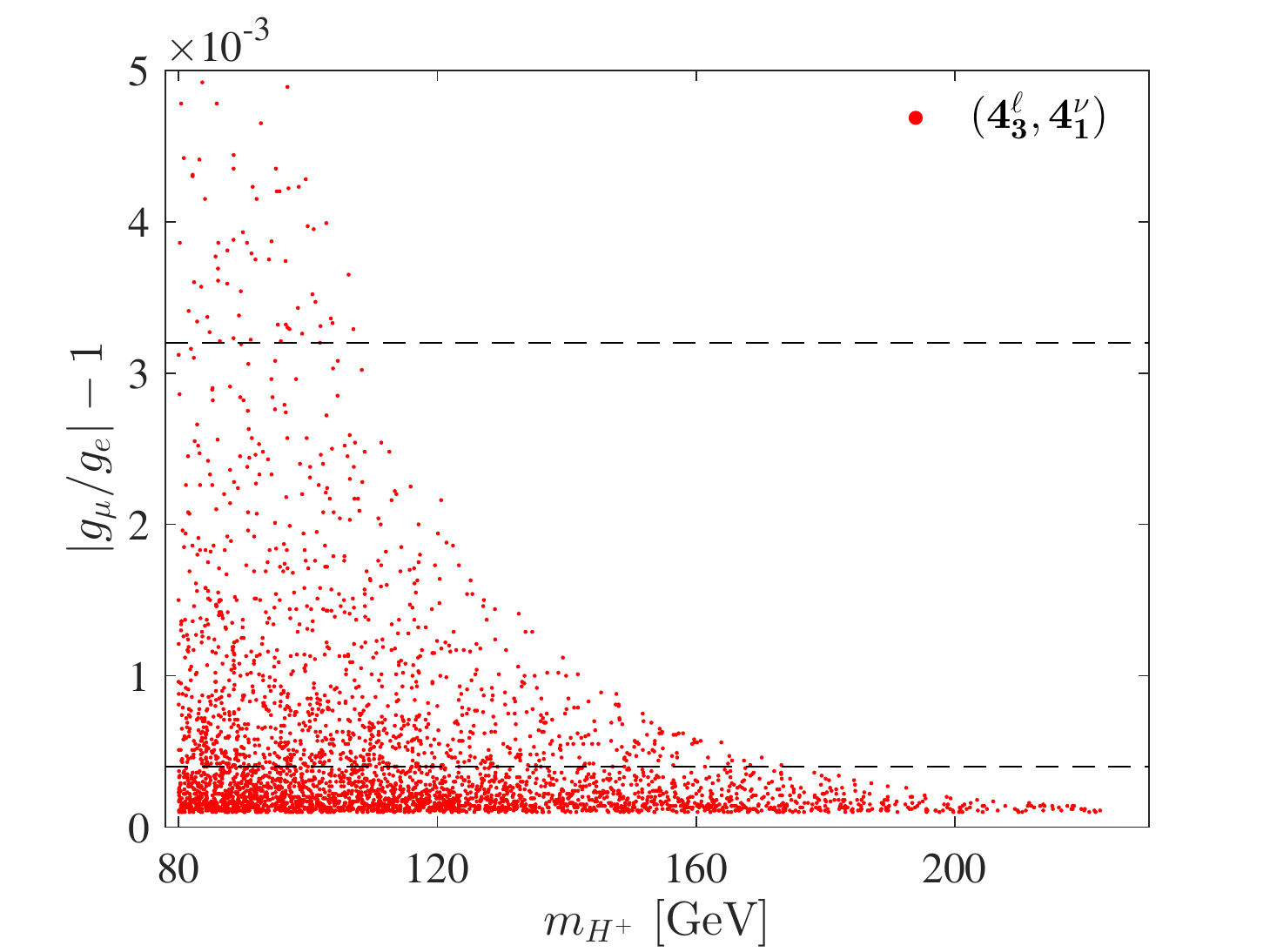} 
		\includegraphics[width=.4\textwidth]{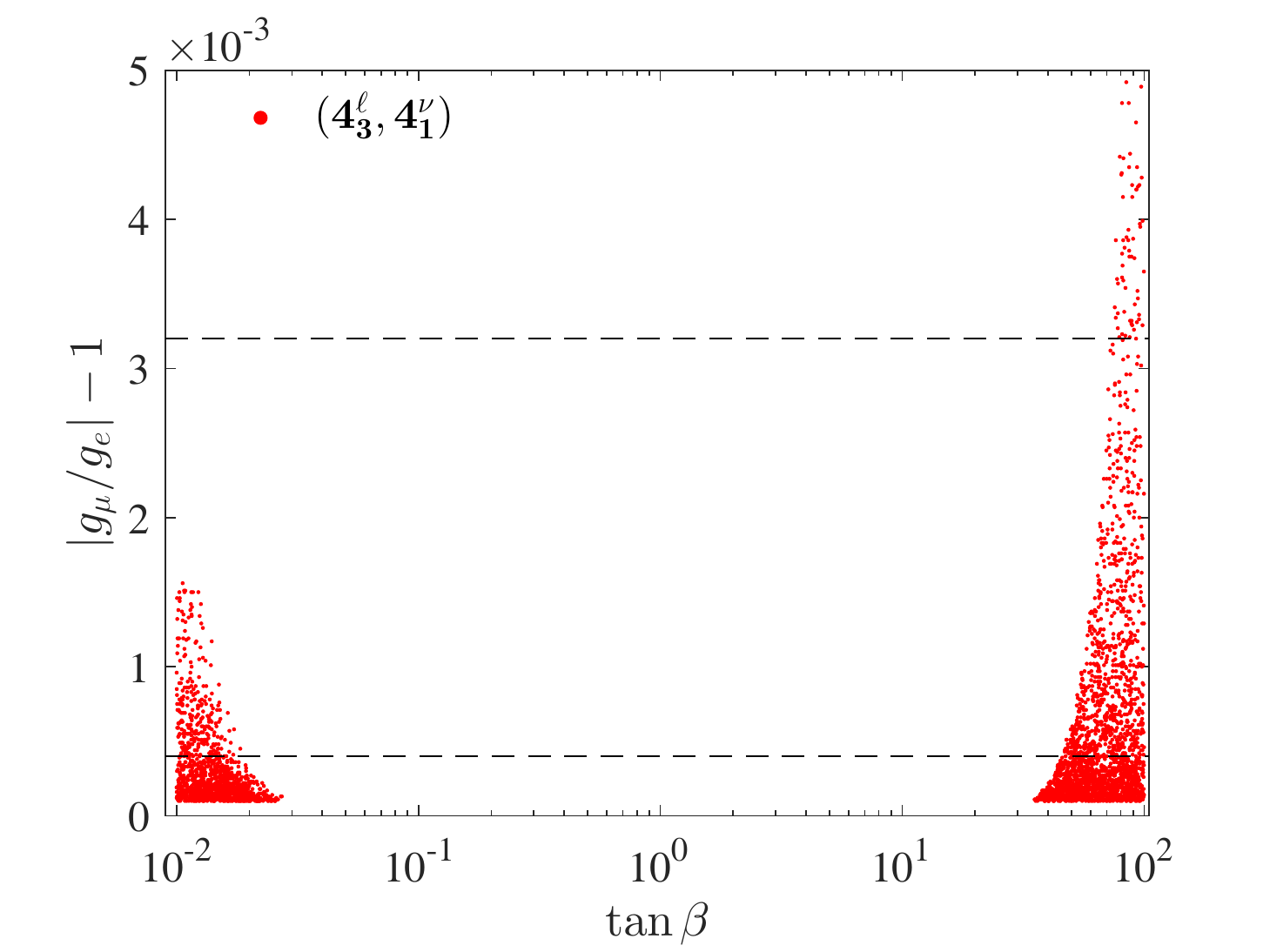}\\ 
		\includegraphics[width=.4\textwidth]{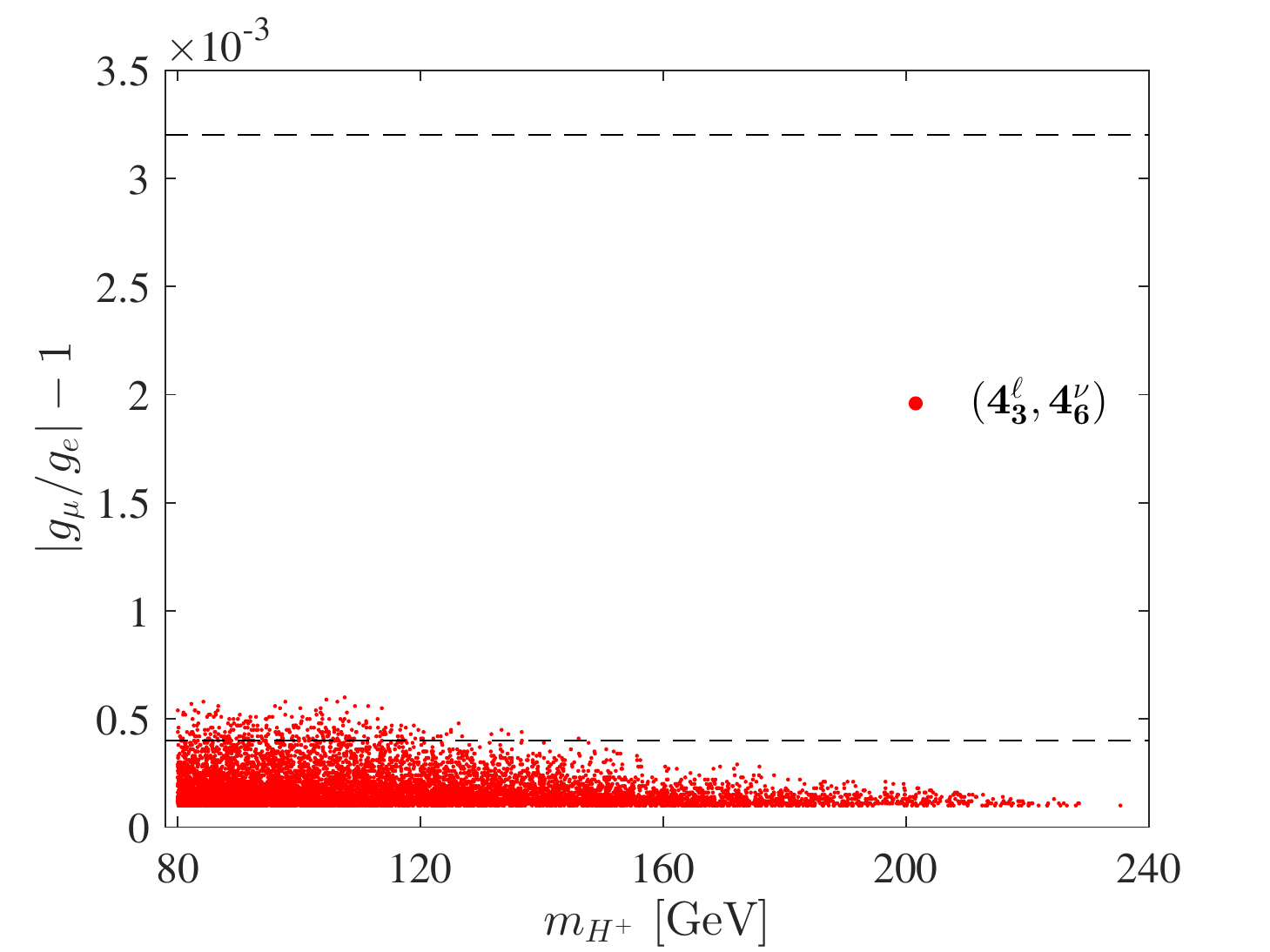}
		\includegraphics[width=.4\textwidth]{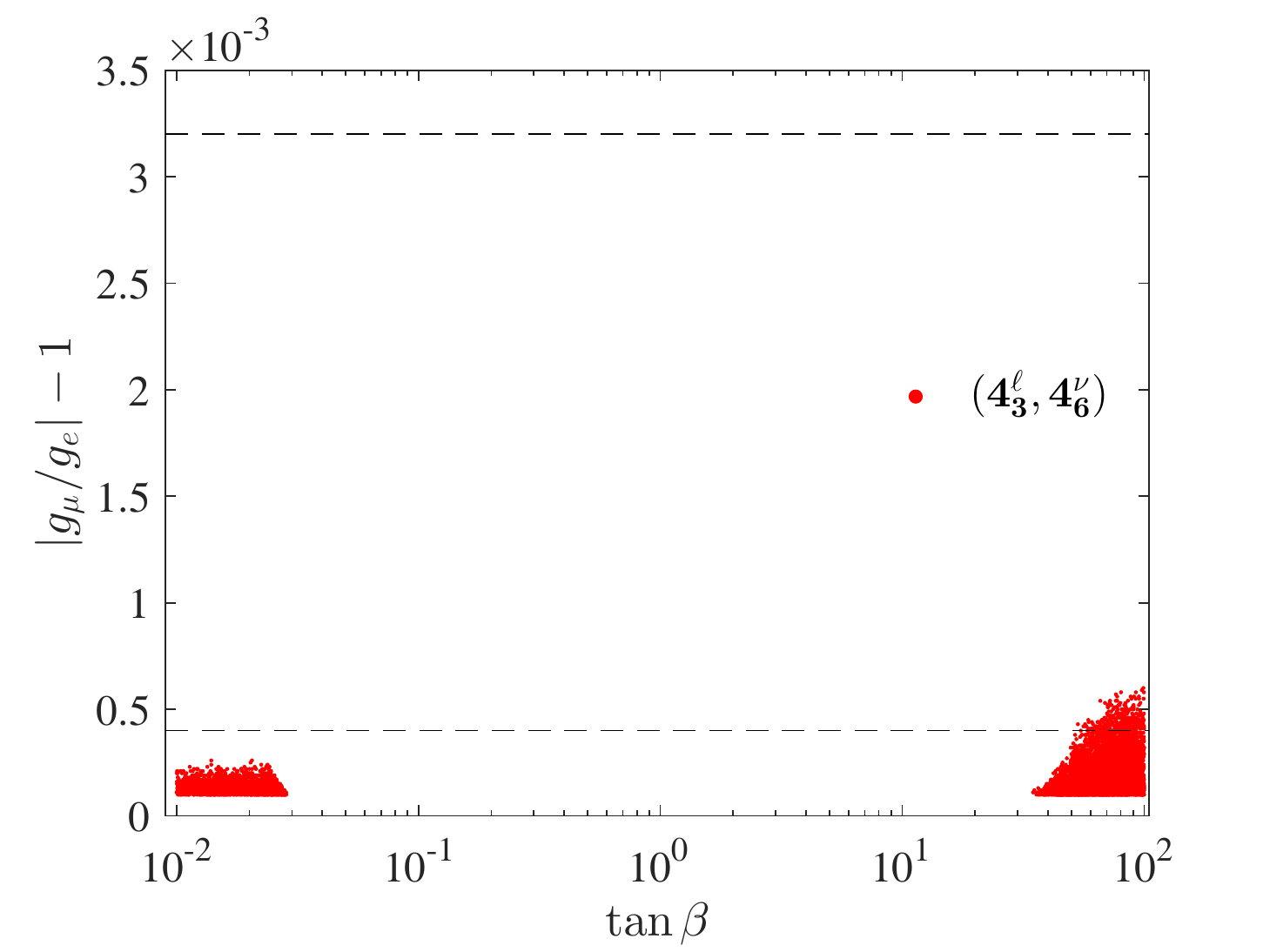}
	\end{tabular}
	\caption{\label{fig1} Charged-lepton universality test in $\tau$ decays for the texture pairs $\big(\mathbf{4_3^{\ell},\mathbf{4_1^{\nu}}}\big)$ and $\big(\mathbf{4_3^{\ell},\mathbf{4_6^{\nu}}}\big)$. The horizontal dashed lines delimit the experimental range given in Eq.~\eqref{eq:expuniv1}. The random points correspond to the parameter space compatible with neutrino data and the bounds of Eqs.~\eqref{eq:expuniv2}, \eqref{eq:expltolll} and~\eqref{eq:expltolgamma}. The deviation from the SM prediction is given as a function of the charged Higgs mass $m_{H^+}$ and $\tan\beta$ in the left and right plots, respectively.}
\end{figure*}

\begin{figure*}[t]
	\centering
	\begin{tabular}{l}
		\includegraphics[width=.4\textwidth]{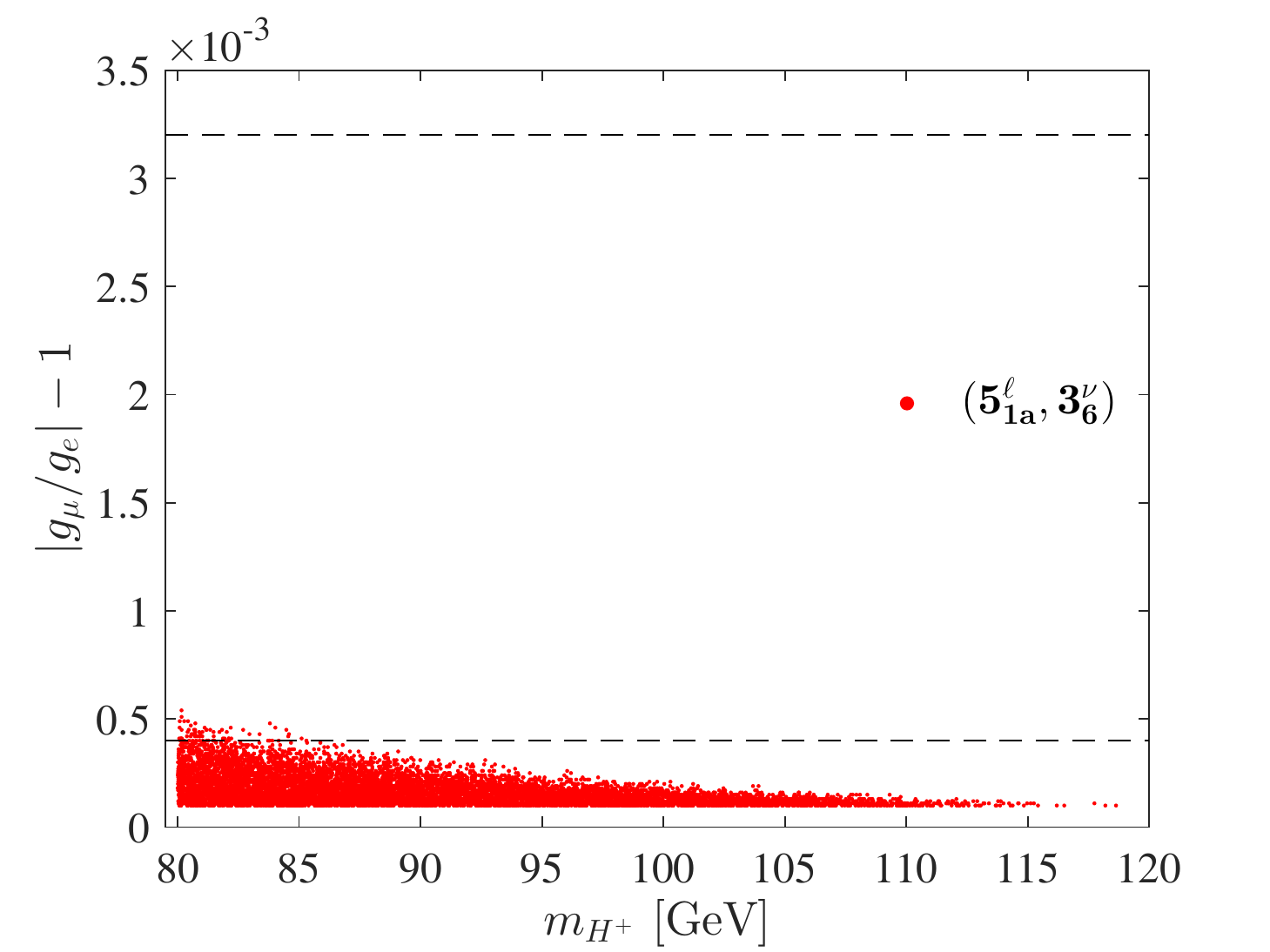} 
		\includegraphics[width=.4\textwidth]{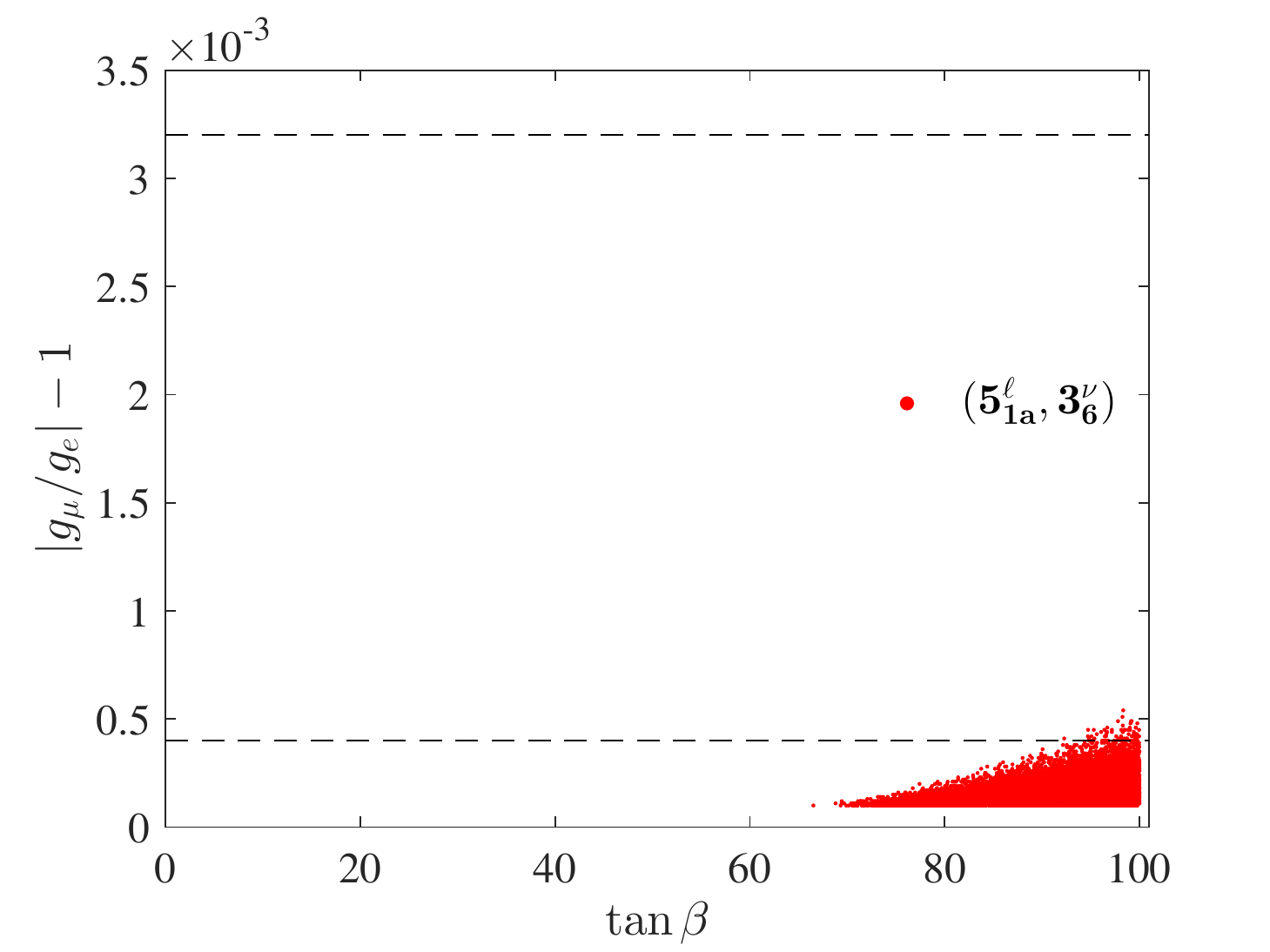}\\ 
		\includegraphics[width=.4\textwidth]{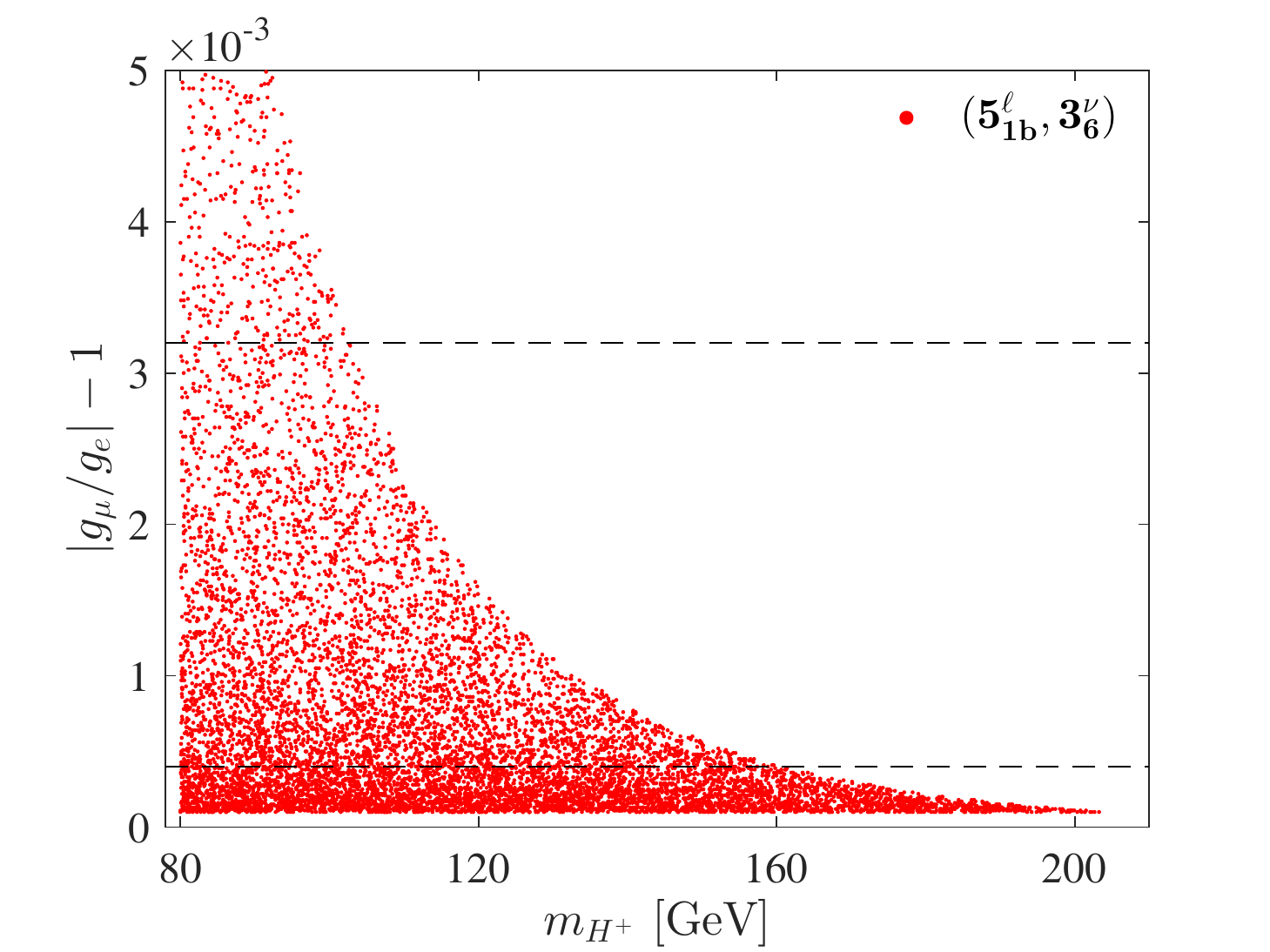}
		\includegraphics[width=.4\textwidth]{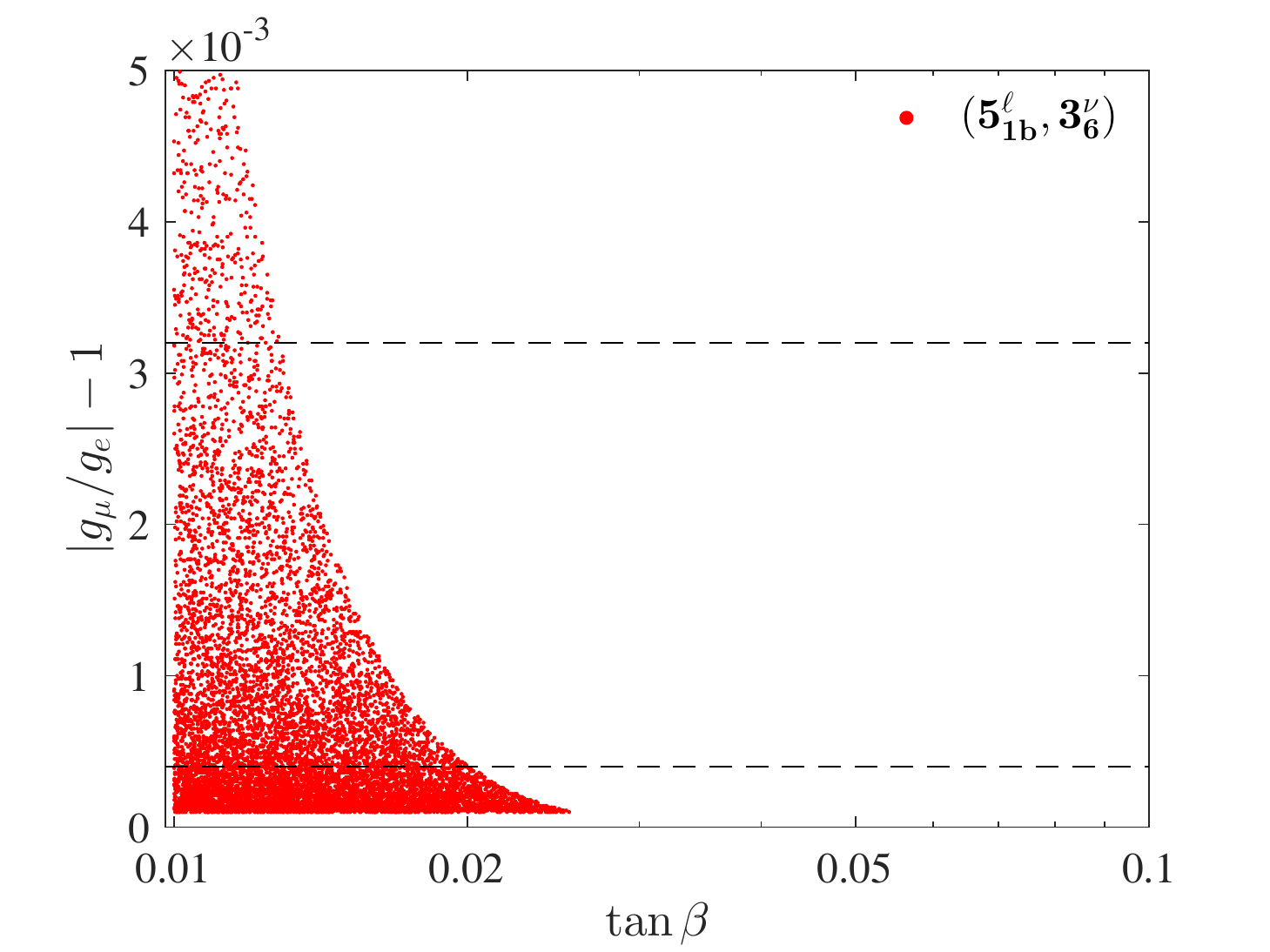}
	\end{tabular}
	\caption{\label{fig2} As in Figure~\ref{fig1}, for the texture pairs $\big(\mathbf{5_{1a}^{\ell},\mathbf{3_6^{\nu}}}\big)$ and $\big(\mathbf{5_{1b}^{\ell},\mathbf{3_6^{\nu}}}\big)$.}
\end{figure*}

\begin{figure*}[t]
	\centering
	\begin{tabular}{l}
		\includegraphics[width=.4\textwidth]{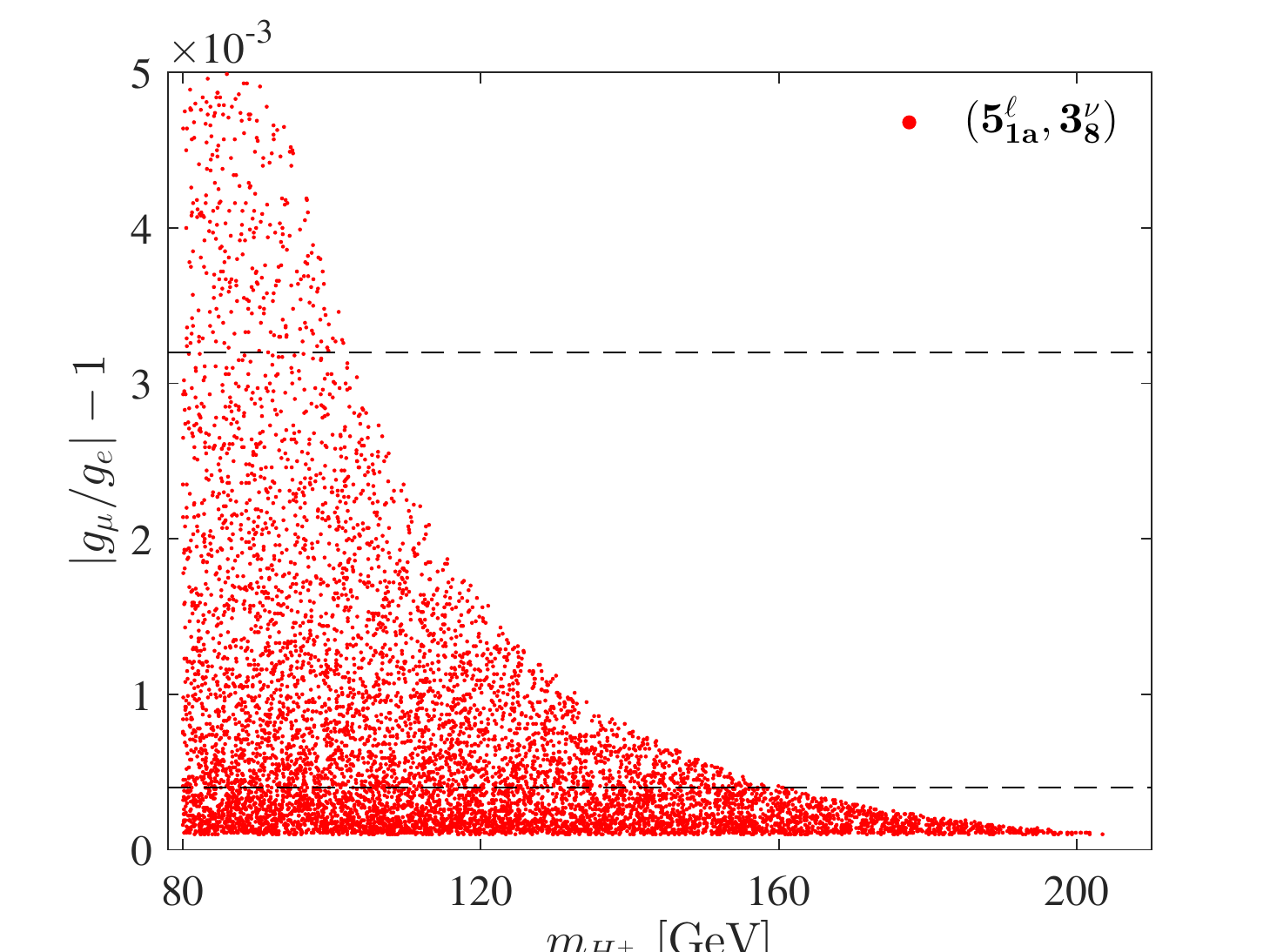} 
		\includegraphics[width=.4\textwidth]{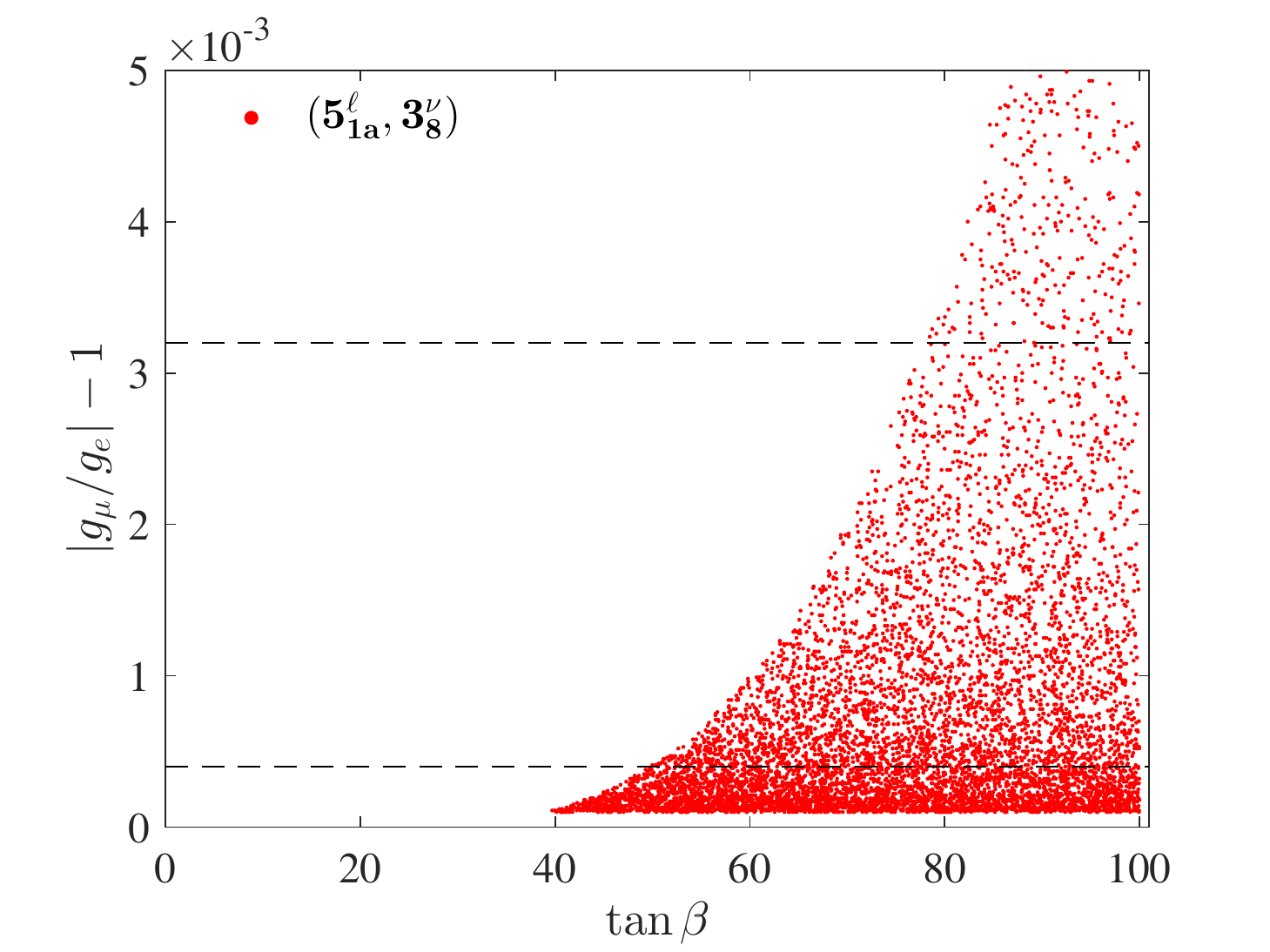}\\ 
		\includegraphics[width=.4\textwidth]{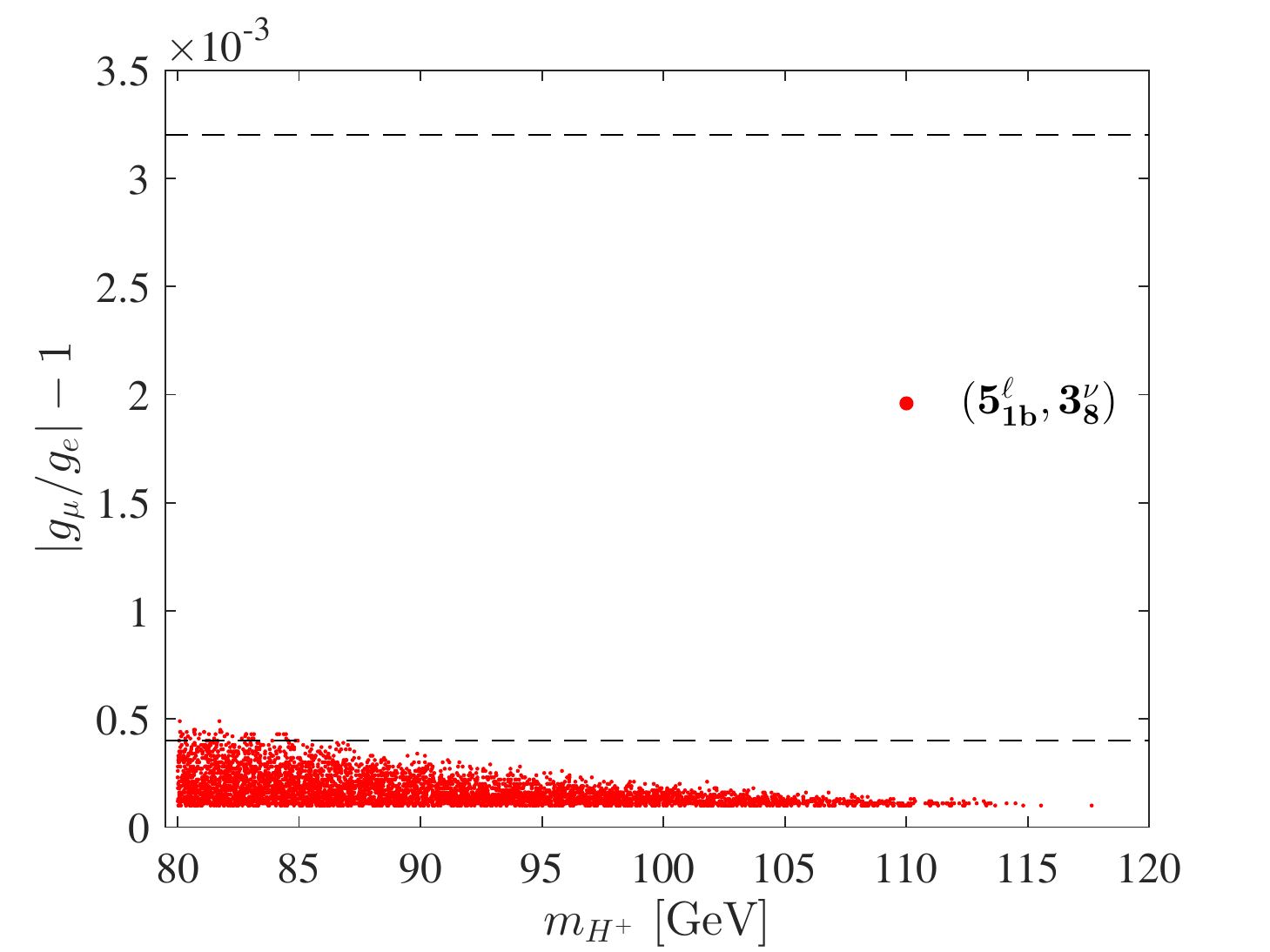}
		\includegraphics[width=.4\textwidth]{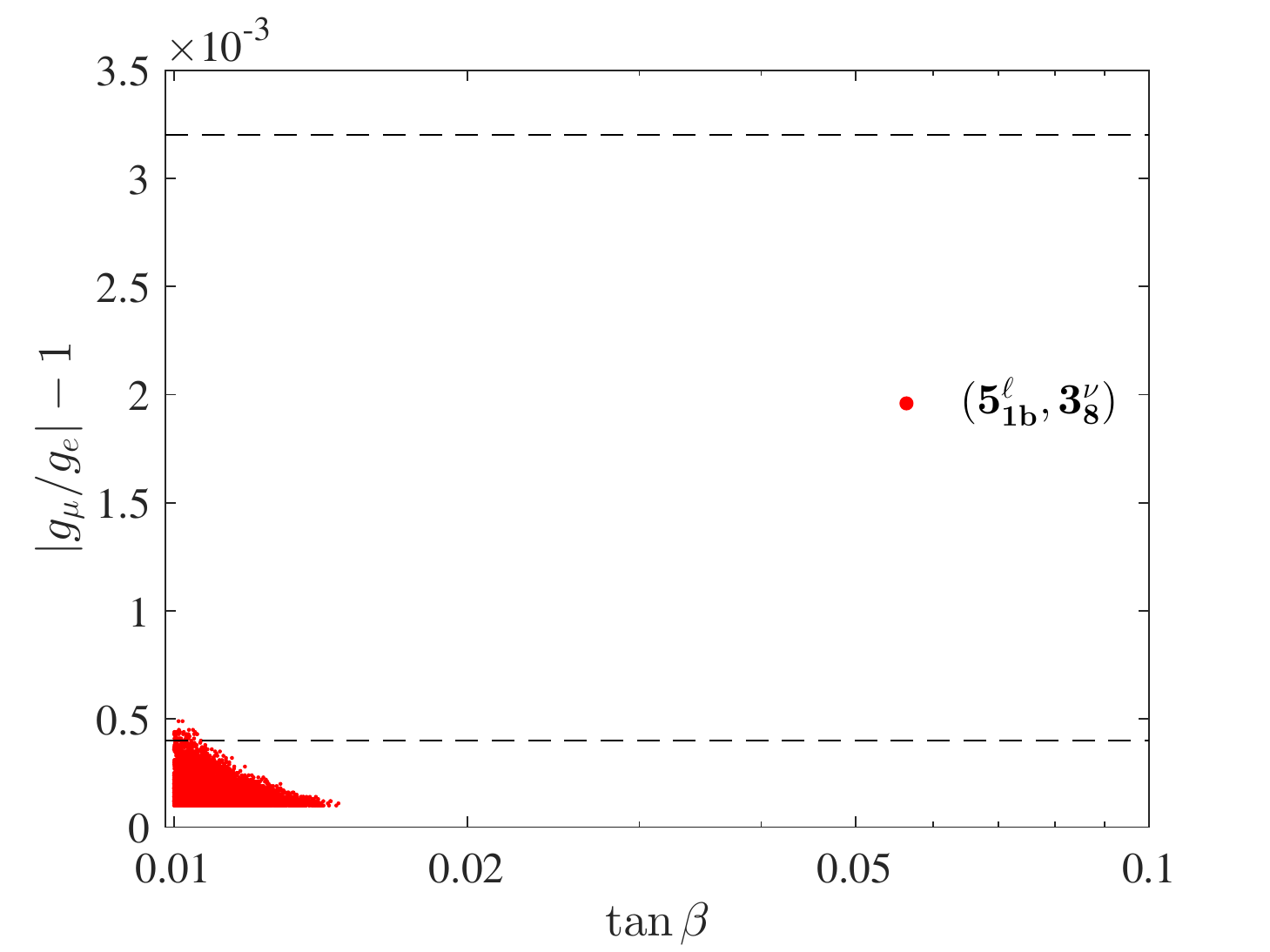}
	\end{tabular}
	\caption{\label{fig3} As in Figure~\ref{fig1}, for the texture pairs $\big(\mathbf{5_{1a}^{\ell},\mathbf{3_8^{\nu}}}\big)$ and $\big(\mathbf{5_{1b}^{\ell},\mathbf{3_8^{\nu}}}\big)$.}
\end{figure*}

\begin{figure*}[t]
	\centering
	\begin{tabular}{l}	
		\includegraphics[width=.4\textwidth]{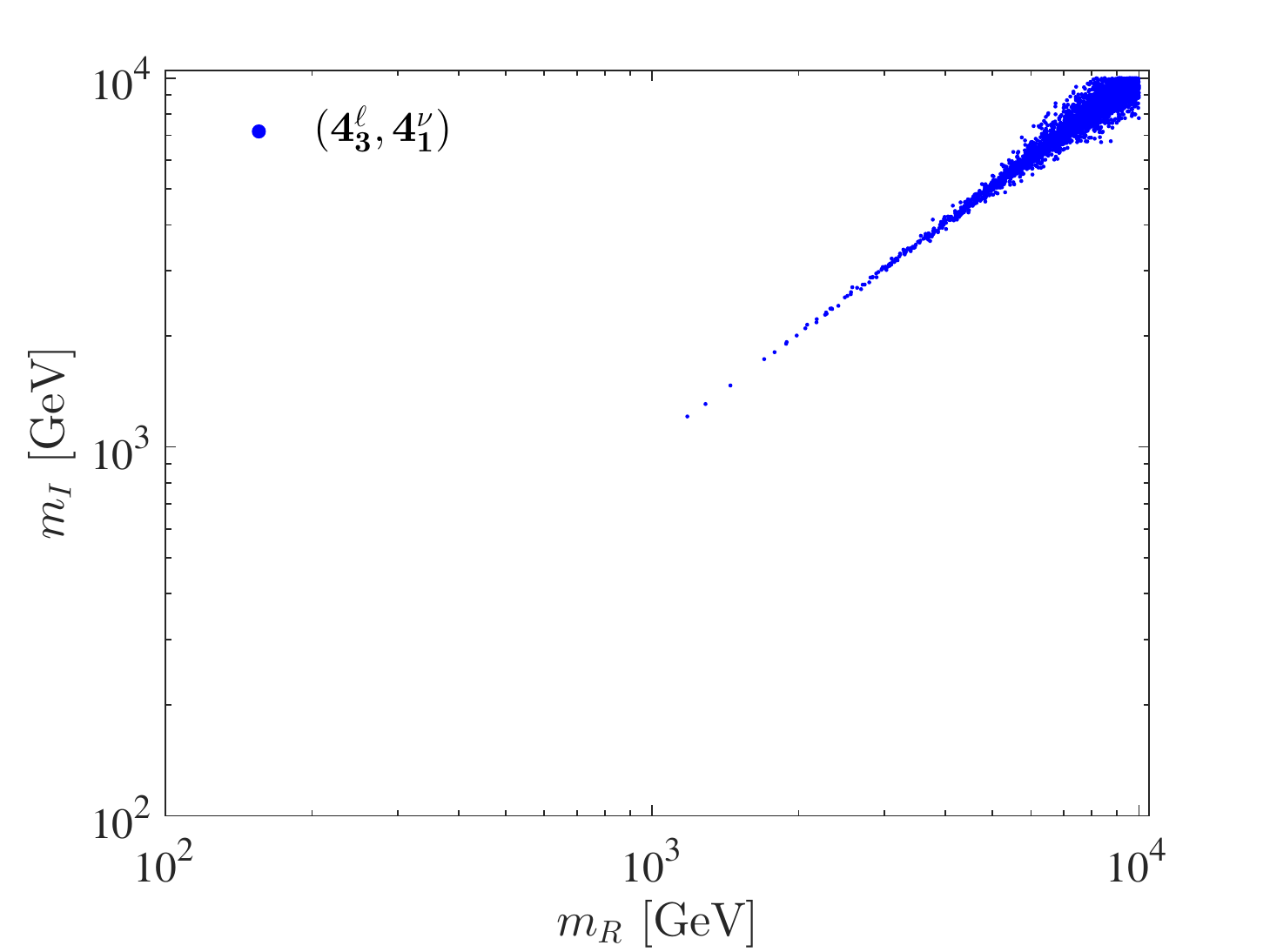} 
		\includegraphics[width=.4\textwidth]{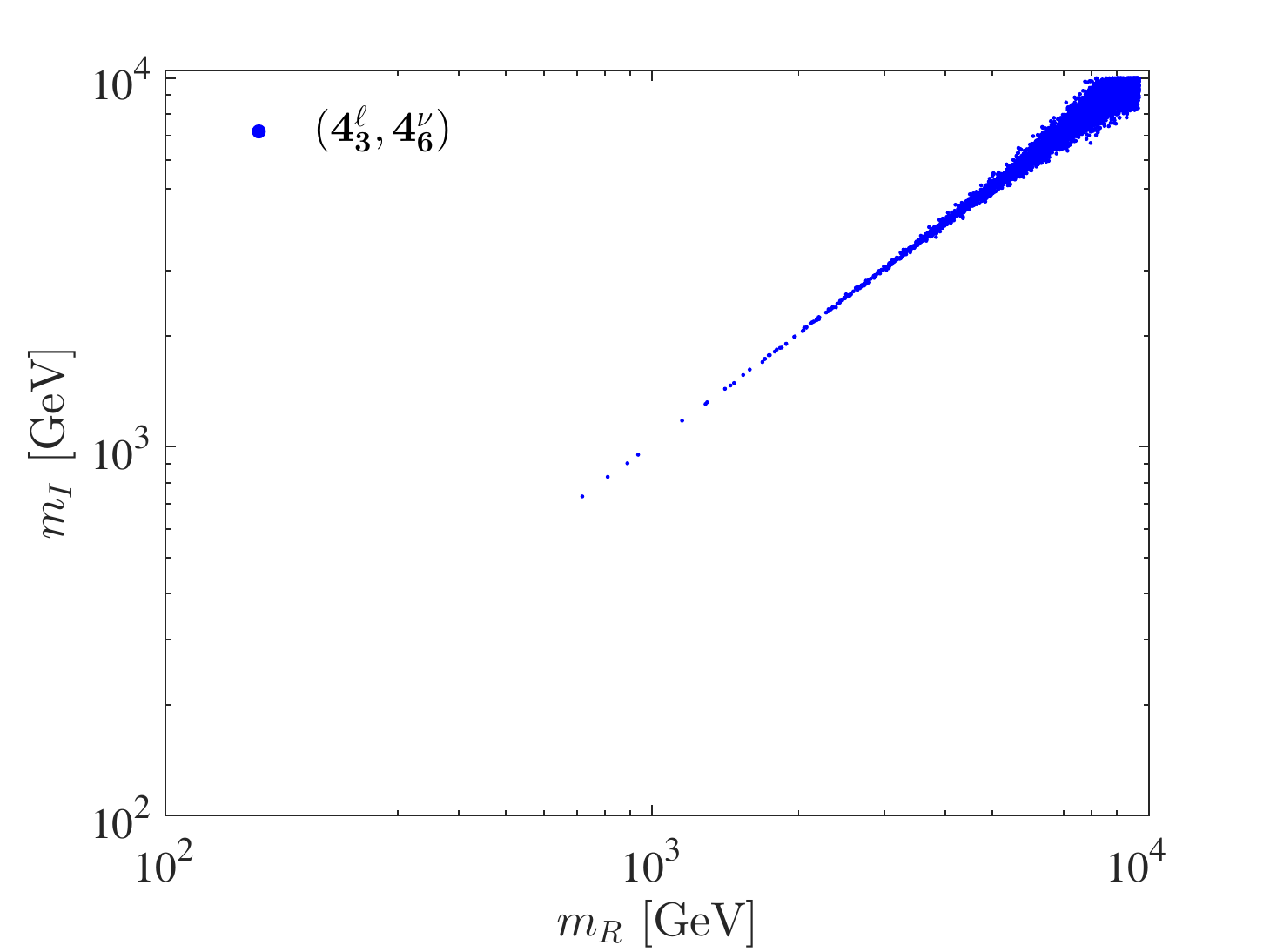}\\
		\includegraphics[width=.4\textwidth]{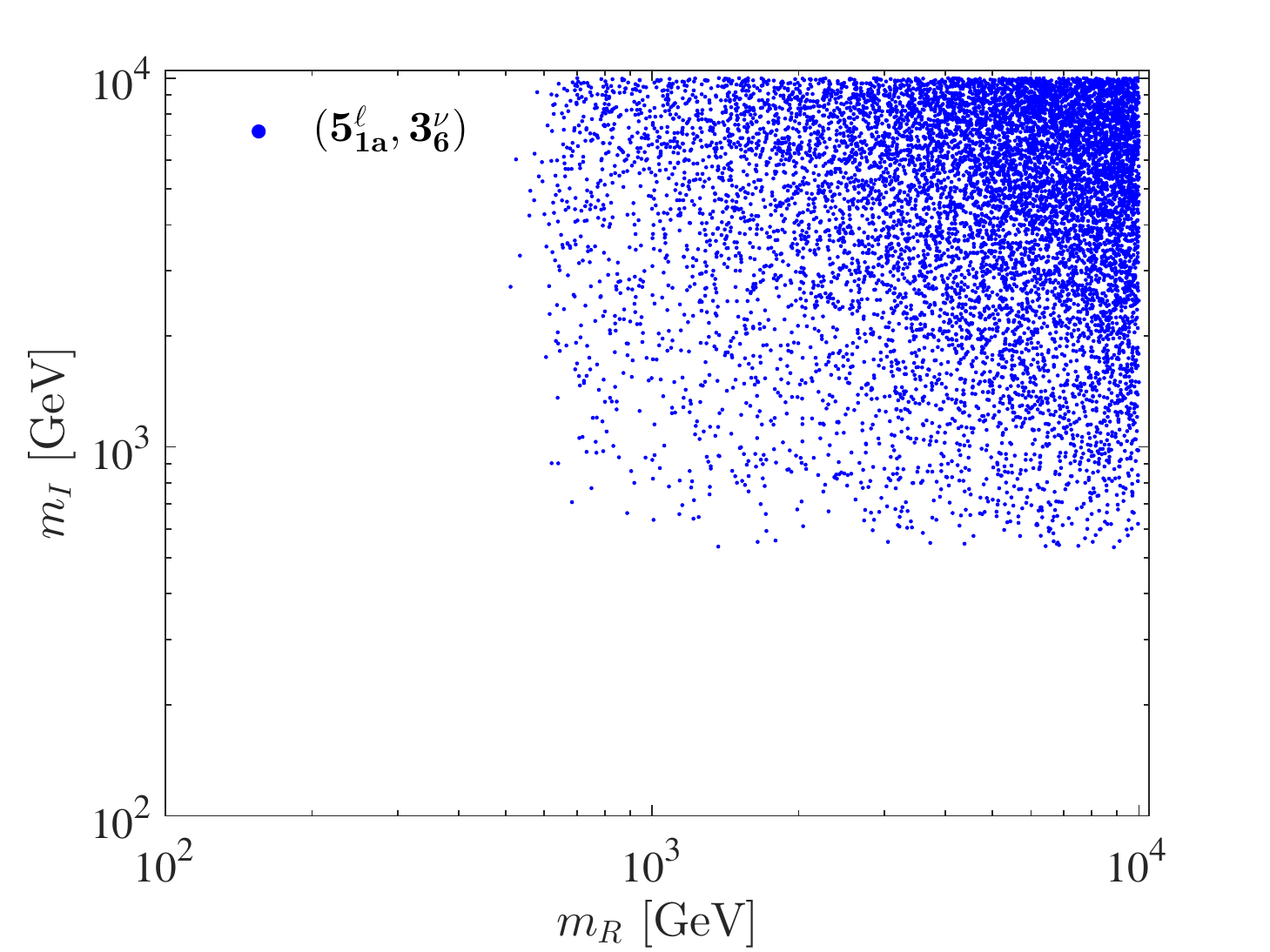}
		\includegraphics[width=.4\textwidth]{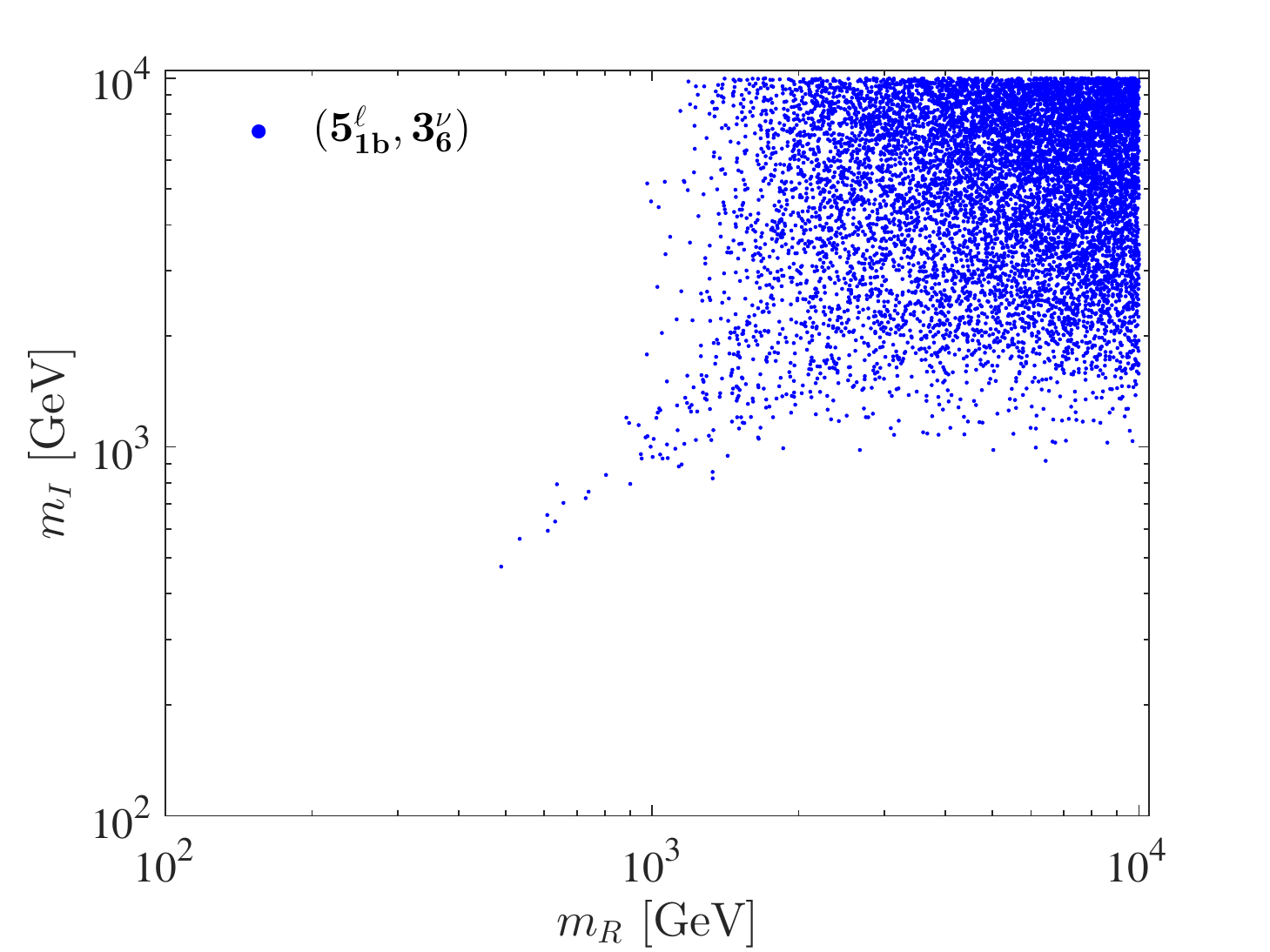}\\
		\includegraphics[width=.4\textwidth]{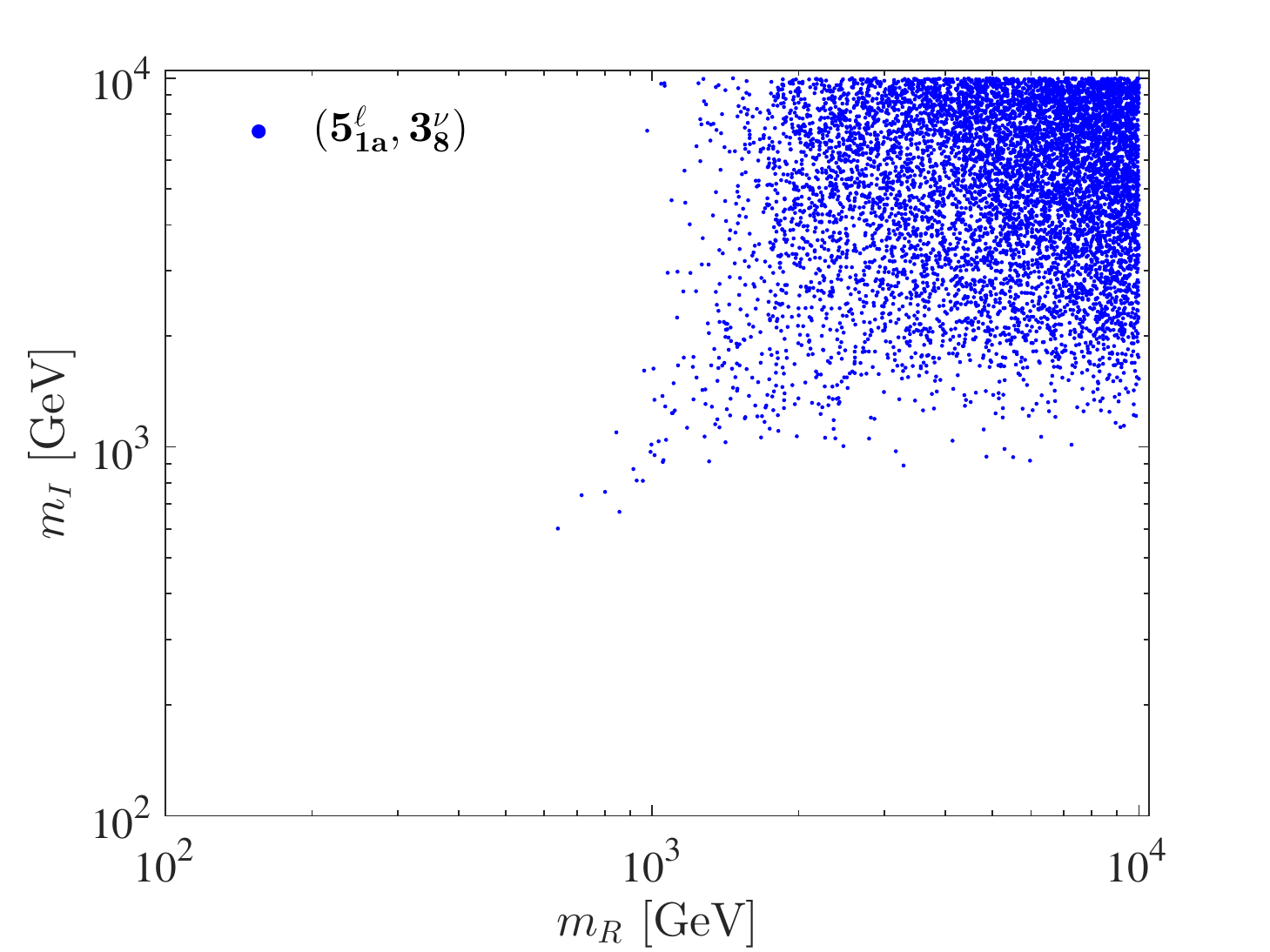}
		\includegraphics[width=.4\textwidth]{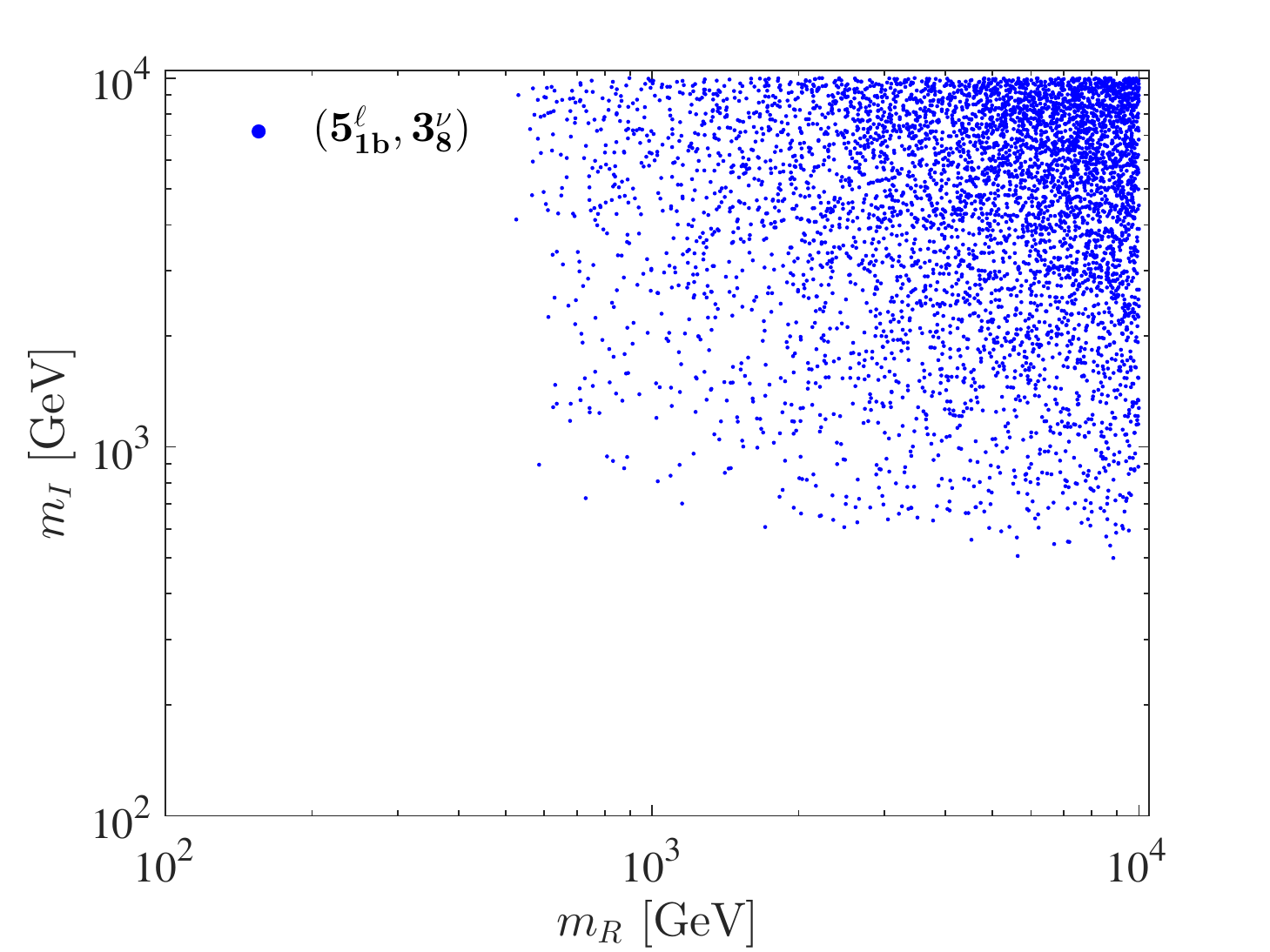}		
	\end{tabular}
	\caption{\label{fig4} The allowed region in the $(m_R,m_I)$-plane for the different charged-lepton and neutrino mass matrix combinations.}
\end{figure*}

\subsection{2HDM: the scalar sector}

Although the detailed study of the scalar potential is beyond the scope of this work, a few comments related to this sector are in order. In the models under consideration, the flavor symmetry restricts the scalar potential in such a way that no explicit CP violation is present in the model. There is no spontaneous CP either and the CP-even and CP-odd scalars do not mix. There remains the mixing in the $(\rho_1,\,\rho_2)$ sector, which can be parametrized through a rotation by a single angle $\alpha$, relating $\rho_1$ and $\rho_2$ to two of the neutral physical Higgs fields,
\begin{align}
\label{eq:physHiggs}
\begin{split}
\begin{pmatrix}
H\\
h
\end{pmatrix}=
\begin{pmatrix}
c_\alpha&s_\alpha\\
-s_\alpha&c_\alpha
\end{pmatrix}
\begin{pmatrix}
\rho_1\\
\rho_2
\end{pmatrix}=
\begin{pmatrix}
c_{\beta-\alpha}&-s_{\beta-\alpha}\\
s_{\beta-\alpha}&c_{\beta-\alpha}
\end{pmatrix}
\begin{pmatrix}
H^0\\
R
\end{pmatrix},
\end{split}
\end{align}
where $h$ and $H$ are the CP-even light and heavy scalars, respectively. Notice also that the last equality in Eq.~\eqref{eq:physHiggs} allows to relate the neutral physical Higgs fields to the neutral real components $H^0$ and $R$ defined in the Higgs basis given by Eqs.~\eqref{eq:Higgstransf} and \eqref{eq:Higgsbasis}.

In the following, we consider the scalar masses independent and we work in the limit of no mixing between $R$ and $H^0$, making the identification of $H^0$ with the Higgs boson discovered by ATLAS and CMS at the LHC~\cite{Aad:2015zhl}, i.e. $m_{H^0}\equiv m_h \simeq 125$~GeV.  The two scalars $R$ and $I$ are then identified with the physical neutral Higgs fields. This limit is obtained by setting $\beta-\alpha=\pi/2$ and it is justified by the SM-like behavior of the observed Higgs particle.

We remark that the presence of a continuous Abelian symmetry would make the CP-odd scalar $I$ a Goldstone boson. The simplest way to deal with such unwanted particle is to softly break the symmetry by introducing, for instance, a mixing term $m_{12}\phi_1^\dagger \phi_2$ in the scalar potential. 

\section{Phenomenological constraints}

It is well known that Yukawa interactions in 2HDM  may induce flavor-changing neutral currents (FCNC) at the tree and loop levels. In this section, we confront the viable maximally restrictive textures of Table~\ref{table4} with current experimental bounds on such processes. In particular, we consider universality in lepton decays mediated by the charged scalar $H^+$ at tree level, the lepton-flavor-violating decays $\ell_\alpha^-\to \ell_{\beta}^-\,\ell_{\gamma}^+\,\ell_{\delta}^-$ mediated by the neutral scalars $R$ and $I$ at tree level and $\ell_\alpha\to \ell_{\beta}\, \gamma$ mediated by neutral and charged Higgs scalars at the one-loop level.\footnote{The phenomenological implications of these processes have been recently studied in Ref.~\cite{Botella:2014ska}, for a class of 2HDM in which the FCNC in the leptonic sector are controlled by the leptonic mixing matrix.}

Before we proceed further, it is worthwhile to comment on the Higgs decays $h\to \ell_\alpha \ell_{\beta}$. Recently, the CMS collaboration, using the data collected in LHC proton-proton collisions at $\sqrt{s}=8$~TeV, reported that no evidence has been found for the lepton-flavor violating decays $h\to e\mu$ and $h\to e\tau$~\cite{Khachatryan:2016rke}. Although an excess in the lepton flavor-violating Higgs decay $h\to \mu\tau$ at a $2.4\sigma$ level has been reported by CMS~\cite{Khachatryan:2015kon}, the ATLAS collaboration observes no evidence for this decay mode~\cite{Aad:2016blu}. Since the decay widths of the above lepton-flavor violating Higgs decays are proportional to the mixing parameter $c_{\beta-\alpha}$, they are naturally suppressed in the decoupling (or alignment) limit of the 2HDM. As for the lepton-flavor conserving decay $h\to \tau \tau$, the signal strength measured at the LHC, normalized to the SM expectation, is so far consistent with the predicted Yukawa coupling strength of the Higgs boson in the SM~\cite{Chatrchyan:2014nva,Aad:2015vsa}.

To study the compatibility of the leptonic textures with current experimental constraints, in our numerical analysis we shall randomly vary the input mass matrices $m_\ell$ and $m_\nu$, the parameter $\tan\beta$ and the Higgs scalar masses $m_{H^\pm}$, $m_R$ and $m_I$. We take $\tan\beta$ as a free parameter varying in the range $10^{-2}$ to $10^2$. As we shall see below, the allowed values of this parameter will depend on the specific model implementation and corresponding experimental bounds.

For the charged Higgs scalar mass, we impose the lower bound $m_{H^\pm}\gtrsim 80$~GeV, obtained from direct searches at colliders~\cite{Abbiendi:2013hk}, while for the neutral Higgs scalars we require $m_{R,I} \gtrsim 100$~GeV~\cite{Agashe:2014kda}. Since we are interested in non-decoupled scenarios, we limit our search to cases where the charged Higgs scalar mass is below 1 TeV and the heavy neutral masses lie in the range up to 10 TeV. 

\subsection{$\ell_\alpha\to \ell_{\beta}\, \nu\, \overline{\nu}$}

Lepton universality tests aim at probing the SM prediction that all leptons couple with the same strength to the charged weak current interaction. Here we shall test this prediction by considering universality in pure $\tau$ decays. The relevant quantity is
\begin{equation}
\label{eq:gmoverge}
\left|\frac{g_\mu}{g_e}\right|^2 \equiv \frac{\text{Br}\left(\tau\to \mu\nu\bar\nu\right)}{\text{Br}\left(\tau \to e\nu\bar\nu\right)}
\frac{f(x_{e\tau})}{f(x_{\mu\tau})}\,,
\end{equation}
where $x_{\alpha\beta}\equiv m^2_\alpha/m^2_\beta$\,. The branching ratios are
\begin{equation}
\label{eq:BRuniv}
\begin{split}
\text{Br}(\ell_\alpha\to \ell_{\beta}\, \nu\,\bar\nu)=
\Big(1+\frac{1}{4}\left|{g_{RR,\alpha \beta}^{S}}\right|^2\Big)
	f(x_{\beta\alpha})+\\
	2\,\text{Re}\left[g_{RR,\alpha \beta}^{S} \left(g_{LL,\alpha \beta}^{V}\right)^\ast\right]
    x_{\beta\alpha}\, g(x_{\beta\alpha})\,,
	\end{split}
\end{equation}
where 
\begin{align}
\begin{split}
f(x)&=1-8x+8x^3-x^4-12x^2 \ln x,\\
g(x)&=1+9x-9x^2-x^3+6x(1+x)\ln x,
\end{split}
\end{align}
are the phase space functions and
\begin{align}
&\left|g_{RR,\alpha \beta}^{S}\right|^2\equiv\sum_{i,j=1}^3|U_{\alpha i}|^2|U_{\beta j}|^2 |g_{i\alpha j\beta}|^2\,,\\
&\left(g_{RR,\alpha\beta}^{S}\right) \left(g_{LL,\alpha \beta}^{V}\right)^\ast\equiv\sum_{i,j=1}^3|U_{\alpha i}|^2|U_{\beta j}|^2 g_{i\alpha j\beta}\,.
\end{align}
Finally, the coefficient $g_{i\alpha j\beta}$ is specific to the model,
\begin{equation}
g_{i\alpha j\beta}=-\frac{(U^\dagger N_e)_{i \alpha} (N_e^\dagger\, U)_{\beta j}}{m_{H^+}^2 U^\ast_{\alpha i} U_{\beta j}},
\end{equation}
with the matrix $N_e$ defined in the basis in which $m_\ell$ is diagonal. 

Current experimental constraints yield~\cite{Amhis:2014hma}
\begin{align}\label{eq:expuniv1}
\left|g_\mu/g_e\right|=1.0018\pm 0.0014
\end{align}
and
\begin{align}\label{eq:expuniv2}
\left|g_{RR,\mu e}^{S}\right|<0.035, \; \left|g_{RR,\tau e}^{S}\right|<0.70, \; \left|g_{RR,\tau \mu}^{S}\right|<0.72,
\end{align}
at 95\% CL~\cite{Agashe:2014kda}.

In the above pure leptonic decays, the exchange of the charged Higgs scalar leads to an amplitude that is parametrized through the effective low-energy coupling $g_{RR,\alpha\beta}^{S} \sim m_\alpha m_\beta/m_{H^+}^2$. Since this coupling is proportional to the lepton masses, the most sensitive decay is $\tau\to \mu\, \bar{\nu}_\mu \nu_\tau$. The bounds given in Eqs.~\eqref{eq:expuniv1} and \eqref{eq:expuniv2} then translate into limits on $\tan\beta$ and $m_{H^+}$. The most stringent limit is obtained from the ratio of the total tau decay widths into the muon and electron modes given in Eq.~\eqref{eq:expuniv1}.

\subsection{$\ell_\alpha^-\to \ell_{\beta}^-\,\ell_{\gamma}^+\,\ell_{\delta}^-$}

The lepton-flavor violating decays $\ell_\alpha^-\to \ell_{\beta}^-\,\ell_{\gamma}^+\,\ell_{\delta}^-$ are mediated by the neutral scalars at tree level. The decay width of these processes is
\begin{align}
\begin{split}
&\Gamma(\ell_\alpha^-\to \ell_{\beta}^-\,\ell_{\gamma}^+\,\ell_{\delta}^-)=\frac{1}{1+\delta_{\beta\delta}}\frac{G_F^2 m_{\alpha}^5}{3\times2^{10}\pi^3}\times\\
&\left[
\left|g_{LL}^{\alpha\beta,\gamma\delta}\right|^2+\left|g_{LL}^{\alpha\delta,\gamma\beta}\right|^2+\left|g_{RR}^{\alpha\beta,\gamma\delta}\right|^2+\left|g_{RR}^{\alpha\delta,\gamma\beta}\right|^2+\right.\\
&\left|g_{LR}^{\alpha\beta,\gamma\delta}\right|^2+\left|g_{LR}^{\alpha\delta,\gamma\beta}\right|^2
+\left|g_{RL}^{\alpha\beta,\gamma\delta}\right|^2+\left|g_{RL}^{\alpha\delta,\gamma\beta}\right|^2-\\
&\left.\text{Re}\left(g_{LL}^{\alpha\beta,\gamma\delta}{g_{LL}^{\alpha\delta,\gamma\beta}}^\ast+g_{RR}^{\alpha\beta,\gamma\delta}{g_{RR}^{\alpha\delta,\gamma\beta}}^\ast\right)
\right]\,,
\end{split}
\end{align}
where
\begin{align}
&g_{LL}^{\alpha\beta,\gamma\delta}=(N_e^\dagger)_{\beta\alpha}(N_e^\dagger)_{\delta\gamma}\left(\frac{1}{m_R^2}-\frac{1}{m_I^2}\right)\,,\\ &
g_{RL}^{\alpha\beta,\gamma\delta}=(N_e)_{\beta\alpha}(N_e^\dagger)_{\delta\gamma}\left(\frac{1}{m_R^2}+\frac{1}{m_I^2}\right)\,,\\
&g_{LR}^{\alpha\beta,\gamma\delta}=(N_e^\dagger)_{\beta\alpha}(N_e)_{\delta\gamma}\left(\frac{1}{m_R^2}+\frac{1}{m_I^2}\right)\,,\\ &
g_{RR}^{\alpha\beta,\gamma\delta}=(N_e)_{\beta\alpha}(N_e)_{\delta\gamma}\left(\frac{1}{m_R^2}-\frac{1}{m_I^2}\right)\,.
\end{align}

Current experimental constraints are
\begin{align}\label{eq:expltolll}
\begin{split}
    \text{Br}(\tau^- \to e^-e^+e^-)&<2.7\times10^{-8},\\
    \text{Br}(\tau^- \to \mu^-\mu^+\mu^-)&<2.1\times 10^{-8},\\ 
    \text{Br}(\tau^- \to e^-\mu^+e^-)&<1.5\times 10^{-8},\\
    \text{Br}(\tau^- \to e^-e^+\mu^-)&<1.8\times 10^{-8},\\ 
    \text{Br}(\tau^- \to \mu^-e^+\mu^-)&<1.7\times10^{-8},\\
    \text{Br}(\tau^- \to \mu^-\mu^+e^-)&<2.7\times10^{-8},\\ 
    \text{Br}(\mu^- \to e^-e^+e^-)&<1.0\times 10^{-12},
\end{split}
\end{align}
at 90\% CL~\cite{Agashe:2014kda}.

We recall that in the SM the amplitudes of the above lepton flavor violating decays are  proportional to the neutrino masses and thus these processes are suppressed. Yet, in the models considered here, there are new contributions mediated by the neutral scalars $R$ and $I$ at tree level, which are now relevant. In the case of muons, the only kinematically allowed decay is $\mu^- \to e^-e^+e^-$. For tau leptons, two types of decays can be distinguished, depending on whether the final state $\ell_{\gamma}^+$ belongs to the same family as one of the negatively charged leptons or not.

\subsection{$\ell_\alpha\to \ell_{\beta}\, \gamma$}

Including two-loop contributions, the decay width of the lepton-flavor violating processes $\ell_\alpha\to \ell_{\beta}\, \gamma$ is given by
\begin{align}
\begin{split}
\Gamma(\ell_\alpha \to \ell_\beta\, \gamma)=&\frac{\alpha\, m_{\alpha}^5 G_F^2}{128\pi^4}\times
\left[\left|\mathcal{A}_L\right|^2+\left|\mathcal{A}_R\right|^2 +\right.\\
&\left.\frac{\alpha^2}{\pi^2}\left(\left|\mathcal{C}\right|^2+
\left|\mathcal{D}\right|^2\right)\right],
\end{split}
\end{align}
where
\begin{align}
\label{eq:AL}
\begin{split}
\mathcal{A}_L=
-\frac{(N_e^{\dagger}U)_{\beta i}(N_{e}^{\dagger}U)_{\alpha i}^\ast}{12\,m_{H^{+}}^2}
+\frac{(N_{e})_{i\beta}^\ast (N_{e})_{i\alpha}}{12}\left(\frac{1}{m_{R}^2}+\frac{1}{m_{I}^2}\right),
\end{split}
\end{align}
\begin{align}
\label{eq:AR}
\begin{split}
\mathcal{A}_R=&
\frac{(N_e)_{\beta i} (N_e)_{\alpha i}^\ast}{12\, m_{R}^2}
-\frac{(N_e)_{\beta i} (N_e)_{i\alpha}m_{i}}{2\,m_{R}^2\,m_{\alpha}}
\left[\frac{3}{2}+\ln\left(\frac{m_{i}^2}{m_{R}^2}\right)\right]+\\
&\frac{(N_e)_{\beta i}(N_e)_{\alpha i}^\ast}{12\, m_I^2}
+\frac{(N_e)_{\beta i}(N_e)_{i \alpha}\,m_{i}}{2\,m_{I}^2\,m_{\alpha}} \left[\frac{3}{2}+\ln\left(\frac{m_{i}^2}{m_{I}^2}\right)\right],
\end{split}
\end{align}
and a sum over $i=e,\mu,\tau$ is implicitly assumed. In Eqs.~\eqref{eq:AL} and \eqref{eq:AR}, contributions proportional to the neutrino masses and subleading terms in $m_\ell^2/m_{R,I}^2$ have been neglected. In leading order, the two-loop terms can be approximated by the expressions 
\begin{align}
\mathcal{C}=\frac{2(N_{e})_{\beta\alpha}(N_{u})_{tt}\,m_t}{m_{R}^2\,m_{\alpha}}
\ln^2\left(\frac{m_t^2}{m_{R}^2}\right),
\end{align}
\begin{align}
\mathcal{D}=\frac{2(N_{e})_{\beta\alpha}(N_{u})_{tt}\,m_t}{m_{I}^2\,m_{\alpha}}
\ln^2\left(\frac{m_t^2}{m_{I}^2}\right),
\end{align}
where $N_u$ is the analogue of the matrix $N_e$ for the up-quark sector, in the basis where the up-quark mass matrix is diagonal. Since the form of $N_u$ depends on the specific model adopted for quarks, we shall not include two-loop contributions in our analysis. Note however that two-loop corrections can dominate over the one-loop contributions in certain cases~\cite{Chang:1993kw}.

Current experimental upper bounds on the branching ratios are
\begin{align}\label{eq:expltolgamma}
\begin{split}
\text{Br}(\mu \to e\gamma)&<5.7\times 10^{-13},\\
\text{Br}(\tau \to e\gamma)&<3.3\times 10^{-8},\\
\text{Br}(\tau \to \mu\gamma)&<4.4\times 10^{-8},
\end{split}
\end{align}
at 90\% CL~\cite{Agashe:2014kda}.

In the SM, the above decay processes are negligible, since their amplitudes are proportional to $m_\nu^2/m_W^2$. On the other hand, in the 2HDM considered here, one expects contributions at the loop level from the flavor-changing couplings of the neutral Higgs scalars $R$ and $I$, which are proportional to $m_\ell^2/m_{R,I}^2$. Processes mediated by the charged Higgs scalar can also be relevant here, since the coupling $H^+ \ell\, \bar{\nu}$ leads to contributions proportional to $m_\ell^2/m_{H^+}^2$.

\subsection{Numerical results}

In Figures~\ref{fig1}-\ref{fig4} we present the allowed regions for the texture pairs $\big(\mathbf{4_3^{\ell},\mathbf{4_1^{\nu}}}\big)$, $\big(\mathbf{4_3^{\ell},\mathbf{4_6^{\nu}}}\big)$, $\big(\mathbf{5_1^{\ell},\mathbf{3_6^{\nu}}}\big)$ and $\big(\mathbf{5_1^{\ell},\mathbf{3_8^{\nu}}}\big)$. The two viable realizations of the charged-lepton mass matrix  $\mathbf{5_{1}^{\ell}}$ are considered, i.e. $\mathbf{5_{1a}^{\ell}}$ and $\mathbf{5_{1b}^{\ell}}$. The random points correspond to solutions compatible with neutrino data and the bounds of Eqs.~\eqref{eq:expuniv2}, \eqref{eq:expltolll} and~\eqref{eq:expltolgamma}.  In all cases, we require that the deviation from universality $|g_\mu/g_e|-1$ is in the range $10^{-4}$ to $5 \times10^{-3}$. This explains why in Figures~\ref{fig1}-\ref{fig3} the charged Higgs mass $m_{H^+}$ does not reach the initially established upper limit of 1 TeV.

It can be seen from the figures that the matrix pairs $\big(\mathbf{4_3^{\ell},\mathbf{4_1^{\nu}}}\big)$ and $\big(\mathbf{4_3^{\ell},\mathbf{4_6^{\nu}}}\big)$ are compatible with data for very small and very large values of $\tan\beta$. On the other hand, the pairs $\big(\mathbf{5_{1a}^{\ell},\mathbf{3_6^{\nu}}}\big)$ and $\big(\mathbf{5_{1a}^{\ell},\mathbf{3_8^{\nu}}}\big)$ favor large values of $\tan\beta$, while the pairs $\big(\mathbf{5_{1b}^{\ell},\mathbf{3_6^{\nu}}}\big)$ and $\big(\mathbf{5_{1b}^{\ell},\mathbf{3_8^{\nu}}}\big)$ are consistent with the bounds for very small values of $\tan\beta$. 

Requiring $|g_\mu/g_e|$ to be in the restrictive range given in Eq.~\eqref{eq:expuniv1}, the charged Higgs scalar mass $m_{H^+}$ is within the range of direct searches at the LHC in all cases. Note also that, if one imposes Eq.~\eqref{eq:expuniv1}, the solutions $\big(\mathbf{4_3^{\ell},\mathbf{4_6^{\nu}}}\big)$, $\big(\mathbf{5_{1a}^{\ell},\mathbf{3_6^{\nu}}}\big)$ and $\big(\mathbf{5_{1b}^{\ell},\mathbf{3_8^{\nu}}}\big)$ are highly disfavored. Furthermore, from Figure~\ref{fig4} we conclude that the solutions $\big(\mathbf{4_3^{\ell},\mathbf{4_1^{\nu}}}\big)$ and $\big(\mathbf{4_3^{\ell},\mathbf{4_6^{\nu}}}\big)$ require quasi-degenerate masses for the neutral Higgs scalars, i.e. $m_R \simeq m_I$. For the textures $\big(\mathbf{5_{1b}^{\ell},\mathbf{3_6^{\nu}}}\big)$ and $\big(\mathbf{5_{1a}^{\ell},\mathbf{3_8^{\nu}}}\big)$, such correlation only exists for neutral scalar masses below 1 TeV, while for the pairs $\big(\mathbf{5_{1a}^{\ell},\mathbf{3_6^{\nu}}}\big)$ and $\big(\mathbf{5_{1b}^{\ell},\mathbf{3_8^{\nu}}}\big)$ no correlation is observed.

Leptonic CP violation is presently an open issue in neutrino physics. Current global oscillation data shows a mild preference for a potentially large value of the Dirac CP phase $\delta$ in the leptonic mixing matrix~\cite{Fogli:2012ua,GonzalezGarcia:2012sz,Forero:2014bxa}. Furthermore, for Majorana neutrinos, there exist two Majorana phases that play a physical role in the neutrinoless double beta ($0\nu\beta\beta$) decay process.\footnote{If the lightest neutrino is massless, only one Majorana phase has physical meaning.} If this decay is mediated by light Majorana neutrinos, the relevant observable quantity is the so-called effective Majorana mass
$m_{\beta\beta}\equiv |\sum_{i} U_{ei}^2 m_i|$.

The maximally restricted leptonic textures considered in this work contain only 8 physical parameters, which allow to fit the eight experimentally known observables, i.e., 3 charged-lepton masses, 2 neutrino mass-squared differences and 3 angles in the leptonic mixing matrix. It is then important to study the predictions of these textures on other low-energy observables such as the leptonic CP-violating phase $\delta$ and the effective Majorana mass $m_{\beta\beta}$. 

\begin{figure}[t]
	\centering
	\begin{tabular}{l}	
		\includegraphics[width=.4\textwidth]{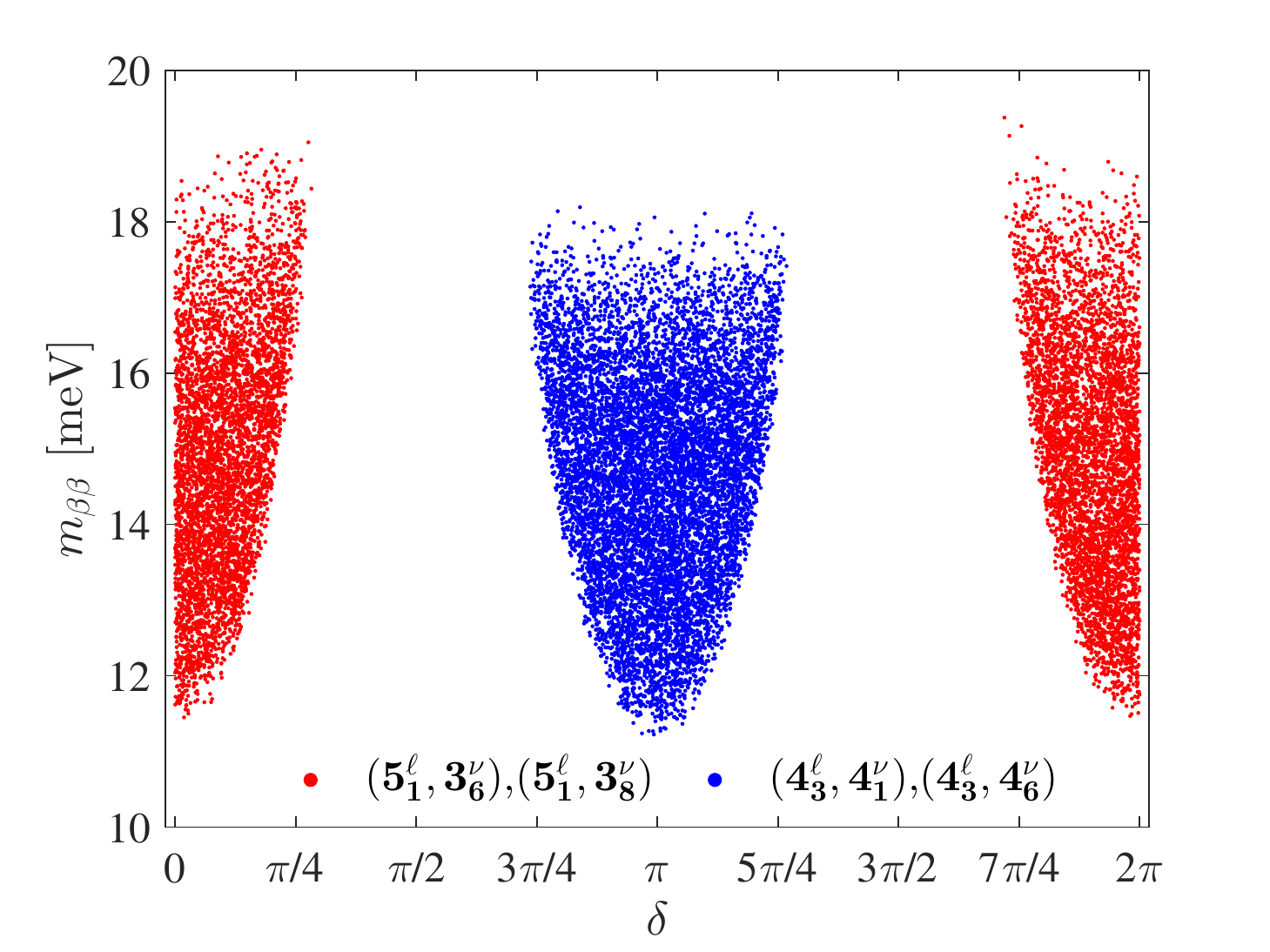}\\
		\includegraphics[width=.4\textwidth]{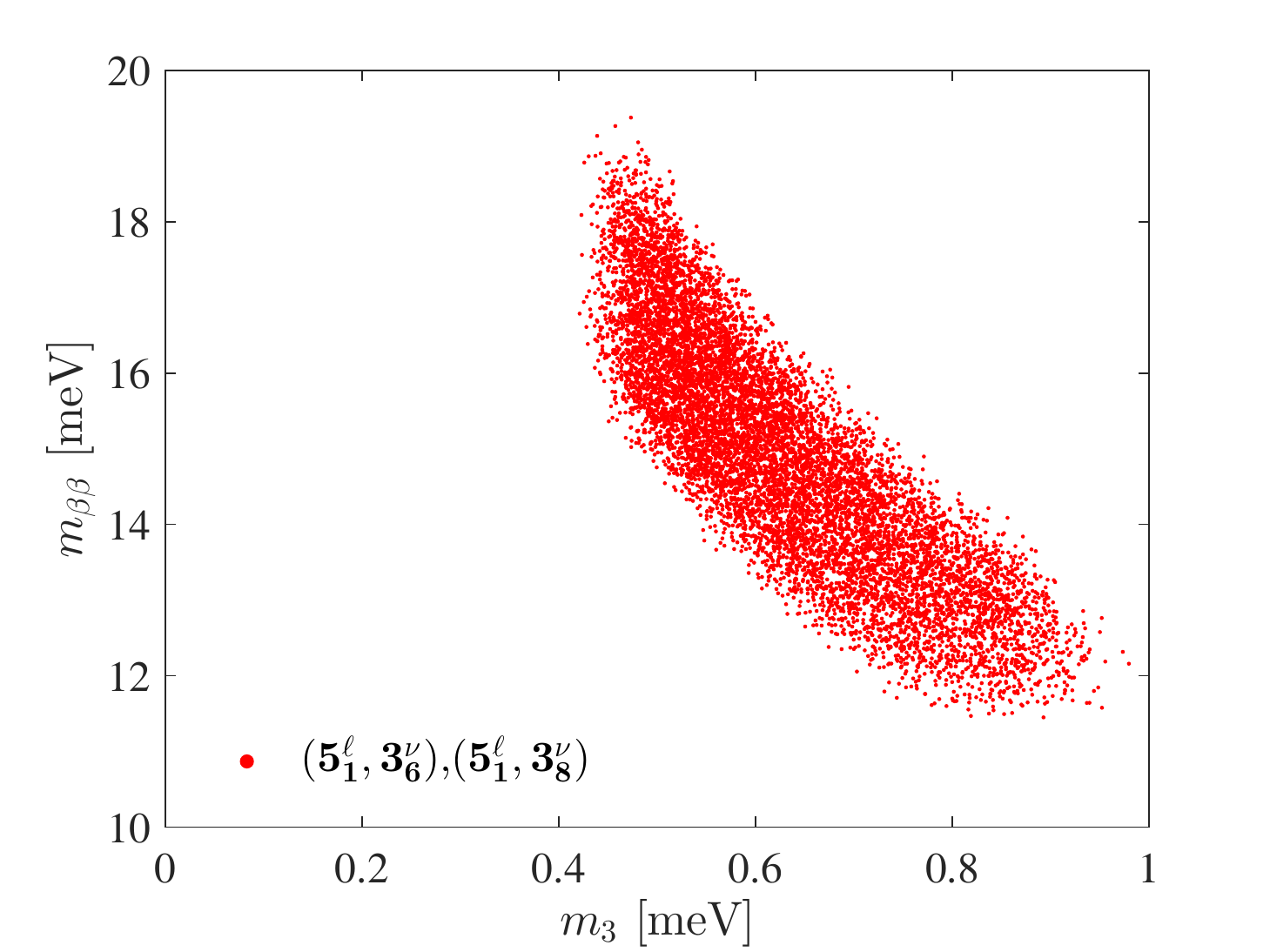}
	\end{tabular}
	\caption{\label{fig5} Predictions of the maximally restrictive texture pairs on the leptonic CP phase $\delta$ and the effective Majorana mass $m_{\beta\beta}$ in $0\nu\beta\beta$ decay. We recall that $m_3=0$ for the pairs $\big(\mathbf{4_3^{\ell},\mathbf{4_1^{\nu}}}\big)$ and $\big(\mathbf{4_3^{\ell},\mathbf{4_6^{\nu}}}\big)$.}
\end{figure}

In Fig.~\ref{fig5}, we present the predicted values for $\delta$ and $m_{\beta\beta}$ for the four viable texture pairs. From the upper panel of the figure we conclude that the allowed region for the CP-violating phase in the cases $\big(\mathbf{5_1^{\ell},\mathbf{3_6^{\nu}}}\big)$ and $\big(\mathbf{5_1^{\ell},\mathbf{3_8^{\nu}}}\big)$ lies approximately in the ranges $(0,\pi/4)$ and $(7\pi/4, 2\pi)$, while for the pairs  $\big(\mathbf{4_3^{\ell},\mathbf{4_1^{\nu}}}\big)$ and $\big(\mathbf{4_3^{\ell},\mathbf{4_6^{\nu}}}\big)$ the allowed range is $(3\pi/4, 5\pi/4)$. No solution compatible with the observations was found around the maximal CP-violating values $\delta=\pi/2$ or $\delta=3\pi/2$. In the lower panel of Fig.~\ref{fig5}, the effective mass parameter $m_{\beta\beta}$ is plotted as a function of the lightest neutrino mass $m_3$ for the pairs $\big(\mathbf{5_1^{\ell},\mathbf{3_6^{\nu}}}\big)$ and $\big(\mathbf{5_1^{\ell},\mathbf{3_8^{\nu}}}\big)$. In the cases $\big(\mathbf{4_3^{\ell},\mathbf{4_1^{\nu}}}\big)$ and $\big(\mathbf{4_3^{\ell},\mathbf{4_6^{\nu}}}\big)$, the lightest neutrino is massless and thus $m_3=0$. The results indicate that the predicted values for $m_{\beta\beta}$ in the four cases are below the expected sensitivity in near future  $0\nu\beta\beta$ experiments~\cite{Alfonso:2015wka,Remoto:2015wta,Dell'Oro:2016dbc}.

\section{UV completion}

The analysis performed so far assumes the existence of the Weinberg operator defined in Eq.~\eqref{Lint}. The dynamics behind this operator has not been specified. As long as the UV completion of the maximally restrictive leptonic textures does not break the effective symmetry, our conclusions remain unchanged. In order to make this statement more definite, we shall present two possible UV completions in the context of type-I and type-II seesaw models.

We shall start with the easiest UV completion of these models, i.e. through the type-II seesaw mechanism. In this framework, several $SU(2)_L$ triplet scalars $\Delta_k$ with hypercharge $Y=1$ are added to the theory,
\begin{equation}
\Delta_k=
\begin{pmatrix}
\Delta_k^0&-\frac{\Delta_k^+}{\sqrt{2}}\\
-\frac{\Delta_k^+}{\sqrt{2}}&\Delta_k^{++}
\end{pmatrix}.
\end{equation}
The relevant Lagrangian terms read
\begin{equation}
-\mathcal{L}_{int}^\text{type-II}=Y_k\overline{\ell_L^0}\Delta_k^\dagger \ell_L^0+\mu_{k,ij}\tilde{\phi}_i^T\Delta_k\tilde{\phi}_j+\text{h.c.},
\end{equation}
where the sum over Latin indexes is implicitly assumed. Once the heavy states with masses $M_k$ are decoupled, the effective coupling matrix reads
\begin{equation}
\frac{\kappa_{ij}}{\Lambda}=-\frac{2\mu_{k,ij}}{M_k^2}Y_k\,.
\end{equation}
The presence of the trilinear couplings $\mu_{k,ij}$ is fundamental in the type-II seesaw, as can be seen in the above formula. For such couplings to be present, $\Delta_k^\dagger$ has to transform under the flavor symmetry as $\tilde{\phi}_i^T\tilde{\phi}_j$. Therefore, if we take as an example the scenario $(\mathbf{5_1^\ell},\mathbf{3_6^\nu})$, we must have
\begin{equation}
\Delta_1\rightarrow e^{i6\gamma}\Delta_1\,,\quad \Delta_2\rightarrow e^{i2\gamma}\Delta_2 \,,\quad \Delta_3\rightarrow e^{i4\gamma}\Delta_3\,.
\end{equation}
The same procedure can be applied for any effective implementation, making the type-II seesaw a trivial path for the UV completion of these models.  

We can also envisage an UV completion within the framework of the type-I seesaw mechanism. However, due to the non-trivial correlation between the UV couplings and the effective neutrino mass matrix, such an implementation may not be viable for all the effective models studied. 

Let us consider, for instance, the effective model $(\mathbf{4_3^\ell}, \mathbf{4_6^\nu})$ with the symmetry transformations given in Eqs.~\eqref{symL1} and~\eqref{symR1b}. We can implement this model in a type-I seesaw framework adding 3 right-handed neutrinos $N_{iR}$ ($i=1,2,3$), which transform appropriately under the flavor symmetry. The requirement that these new fields decouple at low energies implies that the heavy Majorana neutrino mass matrix $M_R$ is non-singular. Together with the constraints at the effective level coming from the $m_\nu$ texture, this leads to the following symmetry transformation:
\begin{equation}
N_{R}\rightarrow \text{diag}\left(e^{-i\gamma},e^{i\gamma},1\right)N_R\,.
\end{equation}
The Lagrangian for the neutrino sector can be written as
\begin{equation}
-\mathcal{L}_{int}^{type-I}=Y_i\overline{\ell_{L}^0}\tilde{\phi}_i N_R+\frac{1}{2}\overline{N_R^c}M_R N_R+\text{h.c.}\,,
\end{equation}
where the Dirac neutrino Yukawa couplings are given by
\begin{equation}
Y_1=
\begin{pmatrix}
0&\times&0\\
\times&0&0\\
0&0&0
\end{pmatrix}\,,\quad
Y_2=
\begin{pmatrix}
\times&0&0\\
0&0&0\\
0&0&0
\end{pmatrix},
\end{equation}
and the heavy neutrino mass matrix is
\begin{equation}
M_R=
\begin{pmatrix}
0&\times&0\\
\times&0&0\\
0&0&\times
\end{pmatrix}.
\end{equation} 
After the decoupling of the heavy neutrinos, we recover the effective Lagrangian in Eq.~\eqref{Lint}, with the coupling $\kappa_{ij}$ predicted by the underlying dynamics,
\begin{equation}
\frac{\kappa_{ij}}{\Lambda}=-Y_iM_{R}^{-1} Y_j^T.
\end{equation}   
From the above textures we can easily verify that $\kappa_{22}$ is the null texture, while $\kappa_{11}$ and $\kappa_{12}$ correspond to the two $\mathbf{4_6^\nu}$ textures found in Eq.~\eqref{46nu}, respectively.

\section{Quark sector}

We now turn our attention to the quarks. Our aim is to implement the same Abelian symmetries in the quark sector without introducing new Higgs scalars. Depending on the charge assignments, different models can emerge, as shown in Table~\ref{table5}. Notice that in the commonly used classification all right-handed fermions of a given charge couple to a single Higgs doublet. Our convention slightly differs from the standard one, since in the models studied here the lepton fields must couple to both Higgs doublets. For the quark fields we choose natural flavor conserving scenarios, i.e. only a single Higgs doublet couples to each quark sector. In particular, models labeled type Ia and type IIa in the table have the same implementation in the quark sector as in the standard type-I and type-II models, respectively. Thus, for such scenarios we can use the current experimental constraints in order to study their viability.

\begin{table}[t]
	\begin{center}
		\begin{tabular}{cccccc}
			\hline\\[-0.3cm]
			Model & $u^0_R$ & $d^0_R$ & $e^0_R$ & $\tan\beta$ & Textures\\[0.1cm]
			\hline\\[-0.3cm]
			Type Ia&$\phi_{2}$ & $\phi_{2}$ & $\phi_{1,2}$ & \multirow{2.5}{*}{large} & \multirow{2}{*}{$\left\{\begin{array}{l}\big(\mathbf{4_3^\ell},\mathbf{4_1^\nu}\big), \big(\mathbf{4_3^\ell},\mathbf{4_6^\nu}\big),\\
			\big(\mathbf{5_{1a}^{\ell},\mathbf{3_6^{\nu}}}\big), \big(\mathbf{5_{1a}^{\ell},\mathbf{3_8^{\nu}}}\big)\end{array}\right.$}\\[0.1cm]
			Type IIa & $\phi_{2}$ & $\phi_{1}$ & $\phi_{1,2}$ & & \\[0.3cm]
			Type Ib & $\phi_{1}$ & $\phi_{1}$ & $\phi_{1,2}$ & \multirow{2.5}{*}{small} &\multirow{2}{*}{$\left\{\begin{array}{l}(\mathbf{4_3^\ell},\mathbf{4_1^\nu}),\\ \big(\mathbf{5_{1b}^{\ell},\mathbf{3_6^{\nu}}}\big), \big(\mathbf{5_{1b}^{\ell},\mathbf{3_8^{\nu}}}\big)\end{array}\right.$}\\[0.1cm]		
			Type IIb & $\phi_{1}$ & $\phi_{2}$ & $\phi_{1,2}$ & & \\[0.1cm]			
			\hline
		\end{tabular}
	\end{center}
	\caption{\label{table5} Model classification according to the coupling of the right-handed quarks $u_R^0, d_R^0,$ and the right-handed charged leptons $e^0_R$ to the Higgs doublets $\phi_{1,2}$. The range for $\tan\beta$ (large or small), required from perturbativity of the quark Yukawa couplings, and the viable pairs of leptonic textures are also presented in each case.}
\end{table}

From Table~\ref{table5}, we conclude that the symmetry implementation in a type-Ia or type-IIa 2HDM framework requires a large $\tan\beta$ to guarantee the perturbativity of the top-quark Yukawa coupling. In both cases, the solutions $\big(\mathbf{4_3^\ell},\mathbf{4_1^\nu}\big), \big(\mathbf{4_3^\ell},\mathbf{4_6^\nu}\big), \big(\mathbf{5_{1a}^{\ell},\mathbf{3_6^{\nu}}}\big), \big(\mathbf{5_{1a}^{\ell},\mathbf{3_8^{\nu}}}\big)$ could be allowed. However, in the type-IIa case the down-quark sector couples only to $\phi_1$ and recent bounds on the weak radiative decay $b\to s\gamma$ imply $m_{H^+} \gtrsim 570$~GeV at 95\% CL~\cite{Misiak:2015xwa,Misiak:2017bgg}. Therefore, the four solutions are excluded in type-IIa scenarios (cf. Figs.~\ref{fig1}-\ref{fig3}).

Despite the above restrictions, the Abelian symmetry can be easily implemented in a type-Ia 2HDM framework. In this case, the bound on $m_{H^+}$ is weaker than the LEP one ($\simeq 80$~GeV) for values of $\tan\beta\gtrsim 2$~\cite{,Misiak:2017bgg}. The quark Yukawa Lagrangian reads
\begin{equation}\label{Lquark}
-\mathcal{L}_{q}^{Y}=\overline{q_L^0}\Gamma d^0_R \phi_2 + \overline{q_L^0}\Delta u^0_R \tilde{\phi}_2+\text{h.c.}\,, 
\end{equation}
where $q_{L}^0$ is the $SU(2)_L$ quark doublet. The Yukawa matrix couplings $\Gamma$ and $\Delta$, associated with the down- and up-quark sectors, respectively, are general $3\times 3$ complex matrices. This implementation simply corresponds to a SM-like Yukawa sector for quarks and, therefore, it is protected from the presence of any FCNC. The symmetry implementation is now quite trivial; we only need to request that $d_{R}^0$ transforms with the opposite charge of $\phi_2$ and $u_R^0$ with the same charge of $\phi_2$. Thus, once the maximally restrictive textures of the leptonic sector are implemented, the Lagrangian terms in Eq.~\eqref{Lquark} can be constructed imposing the quark field transformations
\begin{subequations}
\begin{align}
(\mathbf{4_3^\ell},\mathbf{4_1^\nu}):\quad&d_{R}^0\rightarrow e^{-i3\gamma}d_{R}^0\,,\quad u_{R}^0\rightarrow e^{i3\gamma} u_{R}^0,\\
\nonumber\\
(\mathbf{4_3^\ell},\mathbf{4_6^\nu}):\quad&d_{R}^0\rightarrow e^{i\gamma}d_{R}^0\,,\quad u_{R}^0\rightarrow e^{-i\gamma} u_{R}^0,\\
\nonumber\\
(\mathbf{5_{1a}^\ell},\mathbf{3_6^\nu}):\quad&d_{R}^0\rightarrow e^{-i\gamma}d_{R}^0\,,\quad u_{R}^0\rightarrow e^{i\gamma} u_{R}^0,\\
\nonumber\\
(\mathbf{5_{1a}^\ell},\mathbf{3_8^\nu}):\quad&d_{R}^0\rightarrow e^{i\gamma}d_{R}^0\,,\quad u_{R}^0\rightarrow e^{-i\gamma} u_{R}^0,
\end{align}
\end{subequations}
for each viable texture pair.

\section{Conclusions}

In this paper, we have considered the problem of implementing maximally restrictive texture zeros in the leptonic sector in the context of two-Higgs-doublet models with Majorana neutrinos. We have analyzed all maximally restrictive pairs of leptonic mass matrices with zero entries and concluded that only four pairs of charged-lepton and neutrino mass matrices, namely, $\big(\mathbf{4_3^{\ell},\mathbf{4_1^{\nu}}}\big)$, $\big(\mathbf{4_3^{\ell},\mathbf{4_6^{\nu}}}\big)$, $\big(\mathbf{5_1^{\ell},\mathbf{3_6^{\nu}}}\big)$ and $\big(\mathbf{5_1^{\ell},\mathbf{3_8^{\nu}}}\big)$, are compatible with observations at $3\sigma$ and can be simultaneously implemented through an Abelian symmetry with only  two Higgs doublets. 

We have also investigated whether the viable textures are in agreement with current bounds on lepton-flavor-violating processes. In particular, we have considered the universality in $\tau$ lepton decays mediated by the charged Higgs scalar at tree level, the lepton-flavor-violating decays $\ell_\alpha^-\to \ell_{\beta}^-\,\ell_{\gamma}^+\,\ell_{\delta}^-$ mediated by the neutral Higgs scalars at tree level, and the decay processes $\ell_\alpha\to \ell_{\beta}\, \gamma$ mediated by both neutral and charged Higgs scalars at the one-loop level. The conclusions are summarized in Figures~\ref{fig1}-\ref{fig4}.

Our results turn out to be quite restrictive and predictive. Out of the 19 pairs of maximally restrictive leptonic matrices with 8 physical parameters, only 4 pairs (6 combinations) can be implemented through an Abelian symmetry within the two-Higgs-doublet model. The feasible textures lead to an inverted hierarchical spectrum for the light neutrinos and definite predictions for the Dirac CP-violating observable phase and the effective Majorana mass in $0\nu\beta\beta$ decay, as shown in Fig.~\ref{fig5}. 

By enlarging the number of Higgs doublets we could open the possibility for constructing new pairs of leptonic mass matrices. Nevertheless, we have shown that, in a minimal setup, there are already a few viable scenarios. The latter can be easily completed at the ultraviolet level in the context of the well-known seesaw mechanism through the addition of right-handed singlet neutrinos or the introduction of scalar triplets. The models presented here stem from symmetries and, therefore, they contain a reduced number of free parameters. Once the current experimental constraints are imposed, these models typically lead to definite predictions. 

Finally, we recall once more that the possibility of implementing each set of maximally restrictive textures in the context of 2HDM is a consequence of the existence of an effective Abelian symmetry. Any UV completion respecting this symmetry will clearly lead to the same type of predictions.
 
\section*{Acknowledgements}

We thank F.J. Botella and R. Santos for useful comments and discussions. R.G.F. is grateful to the CERN Theory Division for kind hospitality and acknowledges support from Funda\c{c}\~{a}o para a Ci\^{e}ncia e a Tecnologia (FCT, Portugal) through the grants CERN/FIS-NUC/0010/2015 and UID/FIS/00777/2013. The work of H.S. was partially supported by the Basic Science Research Program through the National Research Foundation of Korea (NRF) funded by the ministry of Education, Science and Technology (No. 2013R1A1A1062597). H.S. has also received funding from the European Research Council (ERC) under the European Union's Horizon 2020 research and innovation programme (grant agreement No. 668679).


\begin{thebibliography}{99}

\bibitem{Fogli:2012ua}
G.~L.~Fogli, E.~Lisi, A.~Marrone, D.~Montanino, A.~Palazzo and A.~M.~Rotunno,
Phys.\ Rev.\ D {\bf 86} (2012) 013012,
doi:10.1103/PhysRevD.86.013012
[arXiv:1205.5254].

\bibitem{GonzalezGarcia:2012sz}
M.~C.~Gonzalez-Garcia, M.~Maltoni, J.~Salvado and T.~Schwetz,
JHEP {\bf 1212} (2012) 123,
doi:10.1007/JHEP12(2012)123
[arXiv:1209.3023].

\bibitem{Forero:2014bxa}
D.~V.~Forero, M.~Tortola and J.~W.~F.~Valle,
Phys.\ Rev.\ D {\bf 90} (2014)  093006,
doi:10.1103/PhysRevD.90.093006
[arXiv:1405.7540].

\bibitem{Strumia:2006db}
  A.~Strumia and F.~Vissani,
  hep-ph/0606054.

\bibitem{Nunokawa:2007qh}
  H.~Nunokawa, S.~J.~Parke and J.~W.~F.~Valle,
  Prog.\ Part.\ Nucl.\ Phys.\  {\bf 60} (2008) 338,
  doi:10.1016/j.ppnp.2007.10.001
  [arXiv:0710.0554].

\bibitem{Branco:2011zb}
  G.~C.~Branco, R.~G.~Felipe and F.~R.~Joaquim,
  Rev.\ Mod.\ Phys.\  {\bf 84} (2012) 515,
  doi:10.1103/RevModPhys.84.515
  [arXiv:1111.5332].

\bibitem{Grimus:2004hf}
  W.~Grimus, A.~S.~Joshipura, L.~Lavoura and M.~Tanimoto,
  Eur.\ Phys.\ J.\ C {\bf 36} (2004) 227
  doi:10.1140/epjc/s2004-01896-y,
  [hep-ph/0405016].

\bibitem{Ivanov:2011ae}
I.~P.~Ivanov, V.~Keus and E.~Vdovin,
J.\ Phys.\ A {\bf 45} (2012) 215201,
doi:10.1088/1751-8113/45/21/215201
[arXiv:1112.1660].

\bibitem{Serodio:2013gka}
H.~Ser\^{o}dio,
Phys.\ Rev.\ D {\bf 88} (2013) 056015,
doi:10.1103/PhysRevD.88.056015
[arXiv:1307.4773].

\bibitem{Ivanov:2013bka}
I.~P.~Ivanov and C.~C.~Nishi,
JHEP {\bf 1311} (2013) 069,
doi:10.1007/JHEP11(2013)069
[arXiv:1309.3682].

\bibitem{Frampton:2002yf}
P.~H.~Frampton, S.~L.~Glashow and D.~Marfatia,
Phys.\ Lett.\ B {\bf 536} (2002) 79,
doi:10.1016/S0370-2693(02)01817-8
[hep-ph/0201008].

\bibitem{Felipe:2014vka}
	R.~Gonz\'{a}lez Felipe and H.~Ser\^{o}dio,
	Nucl.\ Phys.\ B {\bf 886} (2014) 75,
	doi:10.1016/j.nuclphysb.2014.06.015
	[arXiv:1405.4263].

\bibitem{Ludl:2014axa}
	P.~O.~Ludl and W.~Grimus,
	JHEP {\bf 1407} (2014) 090
	[Erratum-ibid.\  {\bf 1410} (2014) 126],
	doi:10.1007/JHEP07(2014)090, 10.1007/JHEP10(2014)126
	[arXiv:1406.3546].

\bibitem{Branco:2011iw}
G.~C.~Branco, P.~M.~Ferreira, L.~Lavoura, M.~N.~Rebelo, M.~Sher and J.~P.~Silva,
Phys.\ Rept.\  {\bf 516} (2012) 1,
doi:10.1016/j.physrep.2012.02.002
[arXiv:1106.0034].
	
\bibitem{Palanque-Delabrouille:2015pga} 
N.~Palanque-Delabrouille {\it et al.},
JCAP {\bf 1511} 011 (2015),
doi:10.1088/1475-7516/2015/11/011
[arXiv:1506.05976].

\bibitem{Alam:2016hwk} 
S.~Alam {\it et al.} [BOSS Collaboration],
arXiv:1607.03155.

\bibitem{Hannestad:2016fog}
S.~Hannestad and T.~Schwetz,
JCAP {\bf 1611} (2016) 035,
doi:10.1088/1475-7516/2016/11/035
[arXiv:1606.04691].

\bibitem{Agashe:2014kda}
K.~A.~Olive {\it et al.}  (Particle Data Group Collaboration),
Chin.\ Phys.\ C {\bf 38} (2014) 090001,
doi:10.1088/1674-1137/38/9/090001.

\bibitem{James:1975dr}
F.~James and M.~Roos,
Comput.\ Phys.\ Commun.\  {\bf 10} (1975) 343,
doi:10.1016/0010-4655(75)90039-9.

\bibitem{Cebola:2015dwa}
L.~M.~Cebola, D.~Emmanuel-Costa and R.~Gonz\'{a}lez~Felipe,
Phys.\ Rev.\ D {\bf 92} (2015) 025005,
doi:10.1103/PhysRevD.92.025005
[arXiv:1504.06594].

\bibitem{Aad:2015zhl}
G.~Aad {\it et al.} [ATLAS and CMS Collaborations],
Phys.\ Rev.\ Lett.\  {\bf 114} (2015) 191803,
doi:10.1103/PhysRevLett.114.191803
[arXiv:1503.07589].

\bibitem{Botella:2014ska}
F.~J.~Botella, G.~C.~Branco, A.~Carmona, M.~Nebot, L.~Pedro and M.~N.~Rebelo,
JHEP {\bf 1407} (2014) 078,
doi:10.1007/JHEP07(2014)078
[arXiv:1401.6147].

\bibitem{Khachatryan:2016rke} 
V.~Khachatryan {\it et al.} [CMS Collaboration],
Phys.\ Lett.\ B {\bf 763} (2016) 472,
doi:10.1016/j.physletb.2016.09.062
[arXiv:1607.03561].

\bibitem{Khachatryan:2015kon} 
V.~Khachatryan {\it et al.} [CMS Collaboration],
Phys.\ Lett.\ B {\bf 749} (2015) 337,
doi:10.1016/j.physletb.2015.07.053
[arXiv:1502.07400].

\bibitem{Aad:2016blu}
G.~Aad {\it et al.} [ATLAS Collaboration],
Eur.\ Phys.\ J.\ C {\bf 77} (2017)  70,
doi:10.1140/epjc/s10052-017-4624-0
[arXiv:1604.07730].

\bibitem{Chatrchyan:2014nva} 
S.~Chatrchyan {\it et al.} [CMS Collaboration],
JHEP {\bf 1405} (2014) 104,
doi:10.1007/JHEP05(2014)104
[arXiv:1401.5041].

\bibitem{Aad:2015vsa} 
G.~Aad {\it et al.} [ATLAS Collaboration],
JHEP {\bf 1504} (2015) 117,
doi:10.1007/JHEP04(2015)117
[arXiv:1501.04943].

\bibitem{Abbiendi:2013hk}
G.~Abbiendi {\it et al.} [ALEPH and DELPHI and L3 and OPAL and LEP Collaborations],
Eur.\ Phys.\ J.\ C {\bf 73} (2013) 2463,
doi:10.1140/epjc/s10052-013-2463-1
[arXiv:1301.6065].

\bibitem{Amhis:2014hma}
Y.~Amhis {\it et al.} [Heavy Flavor Averaging Group (HFAG) Collaboration],
arXiv:1412.7515.

\bibitem{Chang:1993kw}
D.~Chang, W.~S.~Hou and W.~Y.~Keung,
Phys.\ Rev.\ D {\bf 48} (1993) 217,
doi:10.1103/PhysRevD.48.217
[hep-ph/9302267].

\bibitem{Alfonso:2015wka}
K.~Alfonso {\it et al.} [CUORE Collaboration],
Phys.\ Rev.\ Lett.\  {\bf 115} (2015)  102502,
doi:10.1103/PhysRevLett.115.102502
[arXiv:1504.02454].

\bibitem{Remoto:2015wta}
A.~Remoto [NEMO-3 and SuperNEMO Collaborations],
Nucl.\ Part.\ Phys.\ Proc.\  {\bf 265-266} (2015) 67,
doi:10.1016/j.nuclphysbps.2015.06.018

\bibitem{Dell'Oro:2016dbc}
S.~Dell'Oro, S.~Marcocci, M.~Viel and F.~Vissani,
Adv.\ High Energy Phys.\  {\bf 2016} (2016) 2162659,
doi:10.1155/2016/2162659
[arXiv:1601.07512 [hep-ph]].

\bibitem{Misiak:2015xwa}
M.~Misiak {\it et al.},
Phys.\ Rev.\ Lett.\  {\bf 114} (2015) 221801,
doi:10.1103/PhysRevLett.114.221801
[arXiv:1503.01789].

\bibitem{Misiak:2017bgg}
M.~Misiak and M.~Steinhauser,
arXiv:1702.04571 [hep-ph].

\end{thebibliography}
\end{document}